# A global framework to estimate urban spatial cycling patterns based on crowdsourced data

Robert Klein[1], Elias Willberg[1], Silviya Korpilo[2], Tuuli Toivonen[1]

## Abstract

Cycling is a cornerstone of a sustainable mobility transition in cities. Cycling research depends on the data available, but it has been difficult to produce or access these data in comparable ways. Sports tracking platforms like Strava have been transformative in mass-tracking cycling patterns and data sharing through applications and data competitions. Nevertheless, access to data has remained limited. Here, we present a framework that draws on the openly accessible Strava Global Heatmap to estimate spatial patterns of relative cycling intensity on an urban scale. To refine the raw heatmap outputs, we weighted them with population and point of interest (POI) counts within varying buffers. The cycling patterns were validated in a global context, comparing the heatmap values with cycle count data from 29 cities. Both population and POI weighting delivered high correlations in most cases between the heatmap and the cycle counts, POI weighting performing better overall. The strongest associations between Strava heatmap and cycle counts were observed in European cities and along the North American east coast, with $\rho>0.7$ for all, and $\rho>0.8$ for most cities. Additionally, the performance of our approach improved with higher cycling modal share at the city level. We demonstrate that a POI-weighted Strava heatmap can accurately represent urban cycling patterns and provide estimates of categorical cycling volumes. Our approach can be applied with relatively low effort to support the planning for urban cycling if official counts are sparse. Furthermore, it can enable the use of consistent cycling data for large-scale urban cycling analyses.

[1] Digital Geography Lab, Department of Geosciences and Geography, University of Helsinki, Finland
[2] Built Environment Solutions Unit, Finnish Environment Institute, Finland

## 1 Introduction

Cycling offers a range of individual, social, economic, and environmental benefits. Being a comparatively fast mode of transport, cycling can improve accessibility to destinations across populations in urban settings (Jäppinen et al., 2013; Romanillos & Gutiérrez, 2020). Encouraging residents to get on their bikes benefits cities by reducing energy use and pollution from transport systems, supporting climate efforts, and improving air quality (Brand et al., 2021; Cao et al., 2023). Increased cycling levels can also provide significant public health benefits, mainly through higher physical activity and active lifestyles (de Hartog et al., 2010; Oja et al., 2011) as well as reduce traffic congestion (Hamilton & Wichman, 2018; Xu & Zuo, 2024), foster social connectedness (te Brömmelstroet et al., 2017), and save limited urban space (Gössling et al., 2016). As a result, interest in cycling has seen a renaissance in recent decades (Oldenziel et al., 2016). Major cities, such as Copenhagen and Paris, have witnessed substantial increases in modal share in recent years after infrastructure improvements and policy initiatives (Kraus & Koch, 2021; Pucher & Buehler, 2017). While these and some other cities have provided encouraging examples of efforts to strengthen the role of cycling in urban mobility, cycling adoption globally still remains low compared to motorized modes (Goel et al., 2022; Prieto-Curiel & Ospina, 2024).

One of the major barriers for cycling-related planning has long been the lack of spatially accurate, comprehensive, and up-to-date data on cycling volumes available for researchers, planners, and other stakeholders (Aldred et al., 2019). Historically, cycling data collection has often been an afterthought in transport system development, relying on travel diaries, static counters, and manual counts and lacking the systematic counting infrastructure typical of motorized travel. Common challenges in cycling data have included limited spatio-temporal coverage, sporadic data collection cycles, the lack of trip information, underrepresentation of various population groups, and laborious data collection (Bird et al., 2013; Handy et al., 2014). Not knowing where, when, how much, and who among urban residents are using bicycles has hindered efforts to promote cycling, and to develop more cycling-friendly urban environments (Willberg, Tenkanen, et al., 2021).

Recently, advances in geospatial data availability and sensor technologies have started to change the limited cycling data landscape by significantly broadening potential data sources. Cycling activities and behaviors are increasingly recorded and estimated from GPS positioning through various devices and mobile apps (Łukawska et al., 2023; Romanillos et al., 2018; Yan et al., 2025), bike-sharing systems (Levy et al., 2019; O'Brien et al., 2014; Willberg, Salonen, et al., 2021), video recording (Pokorny & Pitera, 2019; Zaki et al., 2013), and street view imagery (Gao & Fang, 2025b). Following the typical shortage of available cycling data, researchers have enthusiastically tapped into the emerging opportunities when estimating, analyzing, and understanding travel patterns of cyclists.

In this respect, the Strava sports application has emerged as a particularly relevant data source for cycling research and planning, quickly becoming an established and widely used source of data. For obtaining these, the company draws on the increasing mass-phenomenon of self-tracking (Lupton, 2016) and assembles data from users who voluntarily track and share their activities and thereby create large amounts of mobility data. Strava serves the community and the broader public by providing segment-level estimates on aggregate cycling volumes with high spatio-temporal resolution across cities worldwide based on its global user population. Even despite its primary focus on sports activities and known socio-economic biases (Venter et al., 2023), Strava has been found to represent activity patterns of cyclists relatively well in various contexts (Lee & Sener, 2021). Access to the Strava Metro platform has been used extensively in urban cycling studies, for example to identify travel patterns (Griffin & Jiao, 2015; Jestico et al., 2016; Selala & Musakwa, 2016), predict cycling demand (Dadashova et al., 2020; Kaiser et al., 2025; Nelson et al., 2021), analyze route choices and traffic safety (Aldred et al., 2018; Orellana & Guerrero, 2019; Sun et al., 2017), understand the impact of infrastructure interventions (Boss et al., 2018; Hong et al., 2020), calibrate travel time estimations (Fink et al., 2024), and estimate the environmental exposure of cyclists (Lee & Sener, 2019). Other researchers have estimated potential noise and errors in the Strava data, and developed techniques to mitigate biases, especially related to socio-demographics (Fischer et al., 2020; Raturi et al., 2021).

Nevertheless, access to the Strava Metro platform has remained restricted, limited to single cities and urban areas, and subject to changes in access policies. This leads to limited opportunities for mobility researchers and practitioners to systematically estimate cycling patterns at an urban scale in high spatio-temporal resolution. As an alternative solution, recent studies have applied Strava Global Heatmap, a visual and interactive representation of aggregated cycling activities collected from Strava users, which is openly available through a dedicated web portal (https://www.strava.com/maps/global-heatmap?sport=Ride). For example, the global heatmap has been applied to estimate urban cycling activities and the presence of people in natural areas (Dong et al., 2023; Gao & Fang, 2025a; Ketchin & Long, 2025). However, to date, no studies have used the global heatmap to collect cycling data across cities and countries, or systematically validated its accuracy against more established cycling data sources.

To fill this gap, the current study presents a novel method for estimating the spatial distribution of cycling intensities across cities at high resolution and urban scale based on the Strava Global Heatmap. We developed a methodological framework for deriving cycling intensities from the heatmap and weighting the derived raw values with points of interest (POI) and population density for more accurate local estimates of cycling levels and distribution. To validate our framework, we compared the cycling patterns we derived with official cycle counts in 29 urban areas in Europe, North and South America, and Oceania across multiple buffer distances. The correlation estimates were further compared with the estimates of cycling modal share, population density, and cycling count density in the study cities. To our knowledge, this is the first study to apply the Strava Global Heatmap to derive cycling patterns across multiple cities in a standardized manner, with the results systematically validated in a global context. The proposed method can serve as a feasible and easily applicable approach to estimate cycling patterns at an urban scale, supporting planners and policymakers to advance sustainable mobility in cities.

## 2 Materials and methods

### 2.1 Research design and study cities

The study design consisted of four parts (Figure 1). First, we developed a methodology to extract spatial estimates of cycling intensities from the Strava heatmap. We then refined these estimates to achieve better local accuracy by weighting them with POI and population density data across multiple buffer sizes. For the third step, we collected official cycling counts and then finally validated the raw and weighted estimates in 29 cities to examine how well they aligned with the cycling counts in each city.

The cities in our study comprised a diverse range of cities across geographic, demographic, and socio-economic contexts and were in multiple countries and continents. From Europe, the cities included were Bern, Berlin, Bordeaux, Copenhagen, Glasgow, Hamburg, Helsinki, London, Lyon, Paris, Valencia, Warsaw, and Zurich. In North America, we included Austin, Bakersfield, Boston, Edmonton, Los Angeles, Montreal, New York, Oakland, Philadelphia, Victoria, and Washington. From South America, we included Buenos Aires and São Paulo, and from Oceania, Christchurch, Perth, and Sydney.

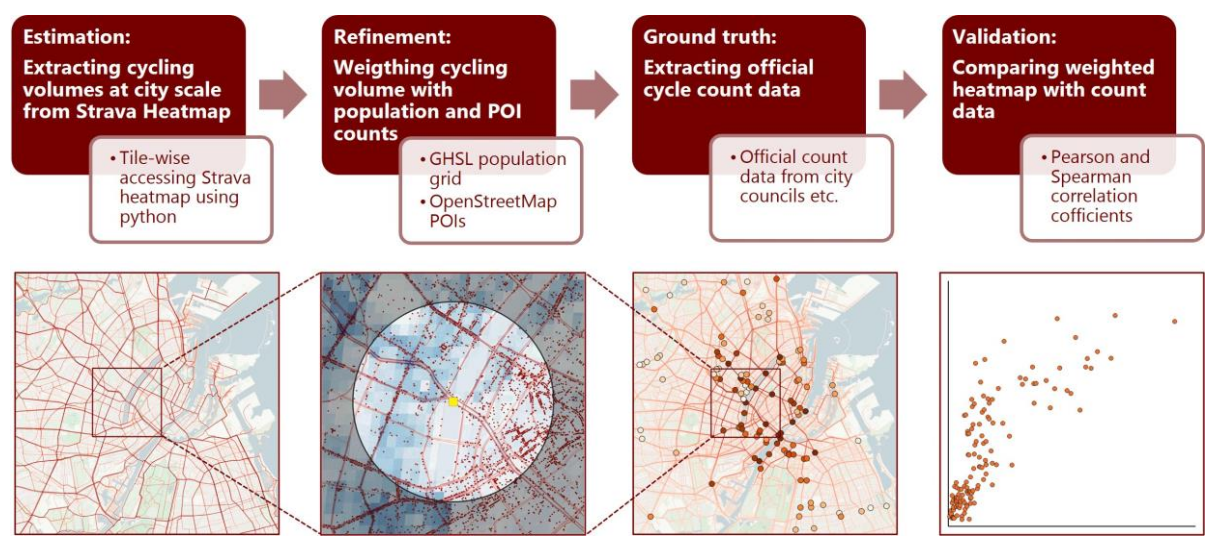

Figure 1: Workflow of the Strava heatmap extraction, refinement, and validation with count data

### 2.2 Data

#### 2.2.1 Strava Heatmap

Strava is an exercise tracking platform on which users can share GPS tracks and metadata of their outdoor activities and compare these activities with others. It has over 150 million worldwide users, and 51 million weekly shared activities make Strava one of the more used platforms of its kind (Strava, 2025a). The shared GPS tracks are collected and aggregated by activity to form the global heatmap, a publicly accessible service indicating relative intensity of exercise at a wide range of locations (Strava, 2025b). The Strava Global Heatmap visualizes aggregate cycling flows in individual road segments across the globe. It considers activities from the previous twelve months and is updated every month.

The global Heatmap is composed of tiles using map tiling in the Web Mercator projection (EPSG:3876), comparable to other popular map services such as Google Maps or OpenStreetMap. Each tile consists of 256x256 pixels with dynamic resolutions depending on the zoom level, with each pixel value representing the amount of shared exercise activities at the given location. For the end-product visualization, the pixel values are calculated using several normalization processes over neighboring tiles, such as histogram equalization, and cumulative distribution functions. Since the heatmap is displayed as a color map, the final pixel values are coded at an ordinal scale 0-255 according to their activity intensity (Robb, 2018).

### 2.2.2 Population data and points of interest

Based on how the Strava heatmap is pre-processed, the raw heatmap values are not comparable over larger areas (Ketchin & Long, 2025). To address this issue and to account for the harmonization and normalization embedded in Strava heatmap (Robb, 2018), we used population and point of interest (POI) distribution as additional inputs to estimate spatial cycling patterns. The underlying assumption is that population and POIs act as proxies for urban activity and would thus influence the number of cyclists in different parts of the city. Both have been previously associated with cycling volumes (Çiriş et al., 2024; Hankey et al., 2021; Zhou et al., 2023). Additionally, POIs and population datasets are openly available globally, and are mapped in a comparable way, which supports the wider applicability of the presented approach. Population data were retrieved from the Global Human Settlement Layer (GHSL). GHSL is a project conducted by the Joint Research Centre of the European Commission that offers a validated, openly accessible global grid containing population estimates at 100 m resolution (GHS-POP R2023A) (Carioli et al., 2023). Additionally, we retrieved POIs from OpenStreetMap (OSM) (OpenStreetMap contributors, 2025) for all the study cities by accessing the raw data from bbbike.org and extracting POIs using the *osmium* Python library. We included a wide variety of POIs but excluded generic tags (e.g., "building") or tags entirely unrelated to cycling (e.g. railway signals). Furthermore, after exploring the data, we excluded POI tags that were very unequally mapped between neighborhoods within cities or between different cities, such as "surveillance" or "streetlamp". This was necessary to ensure better comparability between the cities, and our preliminary tests showed that these exclusions improved the outcomes. A more detailed description of excluded tags is included in Appendix A.

### 2.3 Estimation: Extracting cycling volumes at city scale from Strava Heatmap

To estimate cycling volumes based on the Strava heatmap, we first extracted the raw values. As was done in previous studies (Dong et al., 2023; Gao & Fang, 2025a), we used a python script using Strava account login details to access and georeference the Strava heatmap for cycling ("ride" activities) tile-wise around the counter locations. We used the maximum zoom level of 16 at a spatial resolution of roughly 2.4m. The heatmap was accessed in early May 2025 for the European cities and in early October 2025 for the remaining cities.

## 2.4 Refinement: Weighting cycling volumes with population and POI counts

To account for the pre-processing steps in Strava heatmap, we weighted the raw heatmap output with population density and POI density to examine how well these variables can estimate and represent the actual spatial cycling patterns in the study cities. We created a 100 m grid where we counted for each grid square the population and the number of POIs within 12 buffer sizes (100 m, 200 m, 500 m, 1000 m, 1500 m, 2000 m, 2500 m, 3000 m, 3500 m, 4000 m, 4500 m, and 5000 m). We then weighted the cycling intensity for each heatmap pixel by the respective population and POI counts to generate weighted heatmap rasters.

The weighted heatmap values were calculated as follows:

$$h_{lb,weighted} = h_l \times \frac{n_{lb}}{max\ (n_{bc})}$$

with $h_{lb,weighted}$ as the weighted heatmap value at a given location (l) for a given buffer (b), $h_l$ as the raw heatmap value for the location, $n_{lb}$ as the count of the weighting variable (population or POIs) at the location within the buffer, and $max\ (n_{bc})$ as the maximum count within the buffer in the given city (c).

Additionally, we conducted a combined weighting using both population and POIs at once (pop × POI):

$$h_{lb,weighted} = h_l \times \sqrt{\frac{n\_pop_{lb}}{max\_pop\ (n_{bc})} \times \frac{n\_poi_{lb}}{max\_poi\ (n_{bc})}}$$

2.5 Ground truth: Extracting official cycle count data

We used official and openly available counts as an indicator of the cycling spatial patterns in each city. The count data were released by the municipalities or counties either through their official channels or through sharing the data on the Ecocounter platform.

We adopted the following criteria for the inclusion of cities to aim for spatially representative count data and temporal alignment with the Strava heatmap:

- Recent counts were available from 2023 to 2025. We synchronized the time of the count with the time range covered by the Strava heatmap wherever possible.
- The dataset contained at least 30 count locations (Table 1), with the relatively small Zurich and Bern being exceptions, so we could also test our approach in smaller cities.
- The count locations were spatially dispersed throughout the city and covered not only major cycleways, but also smaller streets.
- The highest counts were more than ten times higher than the lowest ones.

Table 1: The 29 study cities used for validation with the number of count locations (n).

| EUROPE | n | NORTH AMERICA | n | OTHER REGIONS | n |
|---|---|---|---|---|---|
| Bern | 17 | Austin | 30 | Buenos Aires | 473 |
| Berlin | 8821 | Bakersfield | 54 | Christchurch | 36 |
| Bordeaux | 69 | Boston | 106 | Perth | 70 |
| Copenhagen | 137 | Edmonton | 42 | Sao Paulo | 59 |
| Glasgow | 87 | Los Angeles | 78 | Sydney | 77 |
| Hamburg | 66 | Montreal | 50 | | |
| Helsinki | 37 | New York | 32 | | |
| London | 1447 | Oakland | 36 | | |
| Lyon | 83 | Philadelphia | 53 | | |
| Paris | 92 | Victoria | 39 | | |
| Valencia | 123 | Washington | 50 | | |
| Warsaw | 46 | | | | |
| Zurich | 21 | | | | |

The raw counts included in the study originate from various counting approaches and techniques. Our data include continuous counts from automated counters, manual counts throughout a defined timespan, or official count estimates for the entire road network based on automated counts (Berlin). Count locations with no available data for the given timeframe were removed. In the case of automatic counters, locations with a significant time gap in data were excluded. A description of the cycle counts and their sources for each city can be found in Appendix B.

## 2.6 Validation: Comparing the weighted heatmap with count data

As an indicator for cycling volume at each count location, we calculated the heatmap value around each count location for the raw heatmap as well as the weighted rasters. We used the mean value within a 5 m buffer to account for different cycleway/street widths and inaccuracies in the GPS tracks composing the heatmap. Where necessary, we manually moved the count locations to the exact cycleway location. This was needed especially on large boulevards, where the indicated count location was often marked in the middle of the street, whereas the actual cycleway was located on the side.

We then compared the raw and weighted heatmap values to the cycle counts by calculating Pearson and Spearman correlations between each of the series and for every buffer size to test how accurately the respective heatmap values represent actual cycle counts. The focus here is whether the Strava heatmap can be used to estimate overall spatial tendencies for cycling volumes at urban scale, not on predicting absolute cycle counts for specific locations.

Finally, we compared the resulting correlation coefficients with cycling modal share, population density, and spatial count density in the study cities to explore whether the applicability of our method corresponds with these measures. As the count locations did not always cover the entire administrative boundaries of the study cities, we calculated the density measures within a convex hull around all count locations for each city. Comparable cycling modal shares for the most recent available year (2023) were derived from Google Environmental Insights Explorer (Google Environmental Insights Explorer, 2025; see Appendix C), which uses location history data to estimate transportation activity at city scale on yearly basis.

# 3 Results
## 3.1 Estimation and refinement

After accessing the Strava cycling heatmap from the study cities, we obtained georeferenced rasters at the city scale. We observed visibly distinct spatial cycling patterns for both raw and weighted heatmaps. In the raw heatmap, high intensity values were equally distributed throughout the cities' street and cycleway network. However, we observed that because of weighting with population and POIs, the weighted rasters displayed high cycling traffic intensities primarily in central areas of the cities and along major avenues (Figure 2).

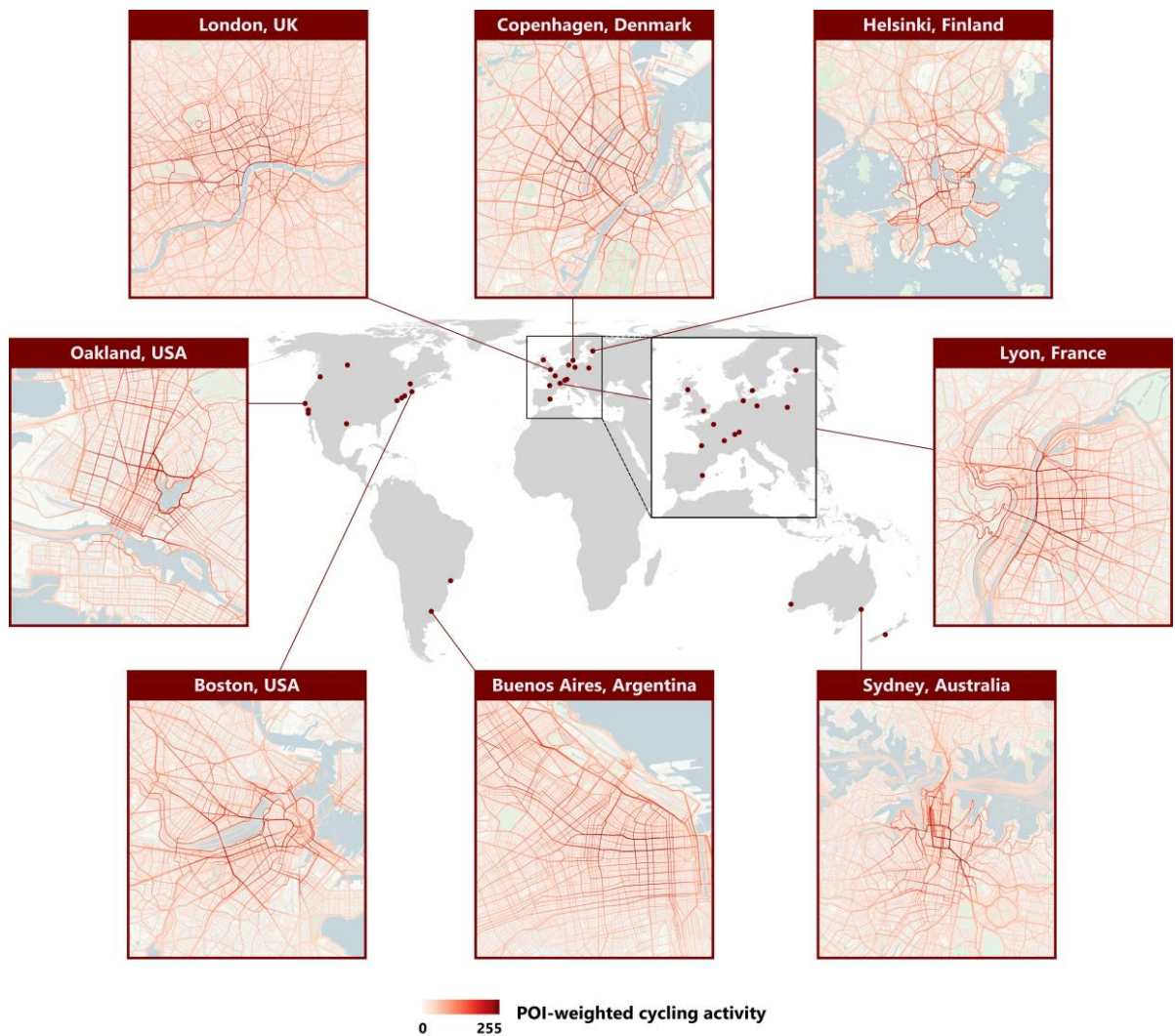

*Figure 2: Overview over the study cities and examples of the obtained POI-weighted rasters*

## 3.2 Validation

For validating our method, we compared the Strava heatmap output with official counts. Even though we calculated both Pearson and Spearman correlations between (weighted) Strava heatmap values and cycle counts, the reported results focus on the Spearman rank correlation coefficients. Rank correlations are more meaningful in this case due to the heatmap using ordinal scales rather than representing absolute cycling volumes and because rank correlations are not assuming linearity. Detailed Pearson correlations are reported in Appendix D.

The validation results indicated that the raw heatmap outputs do not represent spatial cycling patterns well in most of the cities we studied (Figure 3). While moderate relationships ($\rho > 0.6$) between the heatmap and count data existed in single cities, more than half of the study cities exhibited weak correlations ($\rho < 0.4$).

When weighting with population density, correlations between the heatmap and count values increased substantially in Europe, where nearly all study cities displayed strong correlations of $\rho > 0.7$. The highest correlations were found in Lyon ($\rho = 0.86$), Bern ($\rho = 0.85$) and Hamburg ($\rho = 0.84$). In other regions, results were mixed; several cities showed no significant change following population weighting.

Even stronger correlations between heatmap outputs and actual counts were observed when weighting the Strava heatmap with POIs from OSM (Figure 4). Again, we observed the highest correlations in Europe, where all cities exceeded $\rho > 0.7$, with nine out of 13 cities exceeding $\rho > 0.8$. Particularly strong relationships between the POI-weighted Strava heatmap and official counts were found in London ($\rho = 0.89$), Hamburg ($\rho = 0.86$) and Glasgow ($\rho = 0.86$). While this trend was less uniform in other regions, all cities along the North American east coast displayed high correlations ($\rho > 0.7$). Only three cities (Austin, Bakersfield, and Christchurch) remained relatively weakly correlated ($\rho < 0.5$). Overall, POI weighting yielded consistently higher correlations than population weighting across most cities, with Warsaw, Montreal, and Victoria as the only exceptions. For the combined weighting (pop × POI) we did not observe substantial changes in the correlations compared to POI weighting only. While deviations occurred with slightly increased or decreased correlations in single cities, the overall pattern remained almost identical. Detailed results for pop × POI weighting are reported in Appendix E.

### EUROPE

| population buffer (m) | Berlin | Bern | Bordeaux | Copenhagen | Glasgow | Hamburg | Helsinki | London | Lyon | Paris | Valencia | Warsaw | Zurich | average |
|---|---|---|---|---|---|---|---|---|---|---|---|---|---|---|
| raw heatmap output | 0.32 | 0.60 | 0.16 | 0.60 | 0.51 | 0.23 | 0.46 | 0.51 | 0.47 | 0.56 | 0.47 | 0.43 | 0.57 | 0.45 |
| 100 | 0.48 | 0.21 | 0.39 | 0.35 | 0.10 | 0.23 | 0.19 | 0.39 | 0.51 | 0.43 | 0.41 | 0.46 | 0.31 | 0.34 |
| 200 | 0.56 | 0.44 | 0.47 | 0.51 | 0.26 | 0.27 | 0.15 | 0.47 | 0.66 | 0.40 | 0.53 | 0.57 | 0.57 | 0.45 |
| 500 | 0.65 | 0.76 | 0.54 | 0.67 | 0.38 | 0.51 | 0.36 | 0.56 | 0.80 | 0.41 | 0.64 | 0.69 | 0.71 | 0.59 |
| 1000 | 0.70 | 0.79 | 0.66 | 0.74 | 0.56 | 0.76 | 0.50 | 0.65 | 0.84 | 0.45 | 0.68 | 0.79 | 0.75 | 0.68 |
| 1500 | 0.73 | 0.85 | 0.73 | 0.77 | 0.65 | 0.81 | 0.59 | 0.70 | 0.85 | 0.51 | 0.66 | 0.77 | 0.79 | 0.72 |
| 2000 | 0.75 | 0.83 | 0.74 | 0.79 | 0.73 | 0.83 | 0.70 | 0.73 | 0.85 | 0.59 | 0.64 | 0.75 | 0.79 | 0.75 |
| 2500 | 0.75 | 0.80 | 0.74 | 0.81 | 0.76 | 0.84 | 0.77 | 0.75 | 0.86 | 0.69 | 0.62 | 0.75 | 0.79 | 0.76 |
| 3000 | 0.76 | 0.79 | 0.75 | 0.81 | 0.79 | 0.83 | 0.77 | 0.77 | 0.86 | 0.74 | 0.60 | 0.72 | 0.80 | 0.77 |
| 3500 | 0.76 | 0.79 | 0.74 | 0.81 | 0.80 | 0.82 | 0.75 | 0.78 | 0.86 | 0.75 | 0.59 | 0.71 | 0.78 | 0.77 |
| 4000 | 0.76 | 0.81 | 0.72 | 0.80 | 0.79 | 0.81 | 0.75 | 0.79 | 0.86 | 0.76 | 0.58 | 0.73 | 0.73 | 0.76 |
| 4500 | 0.76 | 0.80 | 0.71 | 0.79 | 0.79 | 0.80 | 0.72 | 0.79 | 0.86 | 0.75 | 0.56 | 0.72 | 0.67 | 0.75 |
| 5000 | 0.76 | 0.76 | 0.69 | 0.77 | 0.79 | 0.78 | 0.71 | 0.80 | 0.85 | 0.75 | 0.54 | 0.71 | 0.66 | 0.74 |

### NORTH AMERICA

| population buffer (m) | Austin | Bakersfield | Boston | Edmonton | Los Angeles | Montreal | New York | Oakland | Philadelphia | Victoria | Washington | average |
|---|---|---|---|---|---|---|---|---|---|---|---|---|
| raw heatmap output | 0.13 | 0.31 | 0.47 | 0.46 | 0.06 | 0.27 | 0.10 | 0.25 | 0.36 | 0.27 | 0.40 | 0.28 |
| 100 | 0.04 | 0.28 | -0.05 | 0.03 | 0.01 | 0.32 | 0.32 | 0.06 | 0.29 | 0.18 | 0.11 | 0.15 |
| 200 | 0.15 | 0.28 | 0.14 | 0.14 | 0.07 | 0.5 | 0.5 | 0.24 | 0.32 | 0.24 | 0.21 | 0.26 |
| 500 | 0.04 | 0.31 | 0.37 | 0.31 | 0.06 | 0.66 | 0.53 | 0.31 | 0.55 | 0.35 | 0.41 | 0.36 |
| 1000 | 0.01 | 0.33 | 0.52 | 0.31 | 0.15 | 0.76 | 0.29 | 0.30 | 0.43 | 0.45 | 0.55 | 0.38 |
| 1500 | -0.04 | 0.35 | 0.6 | 0.38 | 0.11 | 0.78 | 0.23 | 0.34 | 0.44 | 0.52 | 0.5 | 0.39 |
| 2000 | 0.15 | 0.36 | 0.67 | 0.48 | 0.1 | 0.75 | 0.3 | 0.36 | 0.55 | 0.6 | 0.55 | 0.45 |
| 2500 | 0.2 | 0.38 | 0.69 | 0.54 | 0.12 | 0.73 | 0.33 | 0.37 | 0.59 | 0.62 | 0.61 | 0.48 |
| 3000 | 0.25 | 0.4 | 0.7 | 0.5 | 0.15 | 0.73 | 0.35 | 0.38 | 0.6 | 0.68 | 0.64 | 0.5 |
| 3500 | 0.32 | 0.43 | 0.73 | 0.5 | 0.17 | 0.7 | 0.34 | 0.33 | 0.63 | 0.71 | 0.67 | 0.52 |
| 4000 | 0.29 | 0.43 | 0.71 | 0.54 | 0.19 | 0.71 | 0.41 | 0.32 | 0.61 | 0.74 | 0.69 | 0.53 |
| 4500 | 0.24 | 0.43 | 0.71 | 0.5 | 0.2 | 0.73 | 0.45 | 0.33 | 0.6 | 0.76 | 0.71 | 0.53 |
| 5000 | 0.31 | 0.42 | 0.69 | 0.5 | 0.22 | 0.72 | 0.46 | 0.34 | 0.59 | 0.77 | 0.72 | 0.54 |

### OTHER REGIONS

| population buffer (m) | Buenos Aires | Christchurch | Perth | Sao Paulo | Sydney |
|---|---|---|---|---|---|
| raw heatmap output | 0.66 | 0.27 | 0.28 | 0.3 | 0.65 |
| 100 | 0.77 | -0.25 | 0.23 | 0.04 | 0.27 |
| 200 | 0.8 | -0.19 | 0.15 | 0.05 | 0.33 |
| 500 | 0.82 | -0.25 | 0.21 | 0.1 | 0.41 |
| 1000 | 0.83 | -0.28 | 0.32 | 0.18 | 0.59 |
| 1500 | 0.83 | -0.21 | 0.38 | 0.17 | 0.68 |
| 2000 | 0.82 | -0.07 | 0.41 | 0.17 | 0.74 |
| 2500 | 0.82 | 0.08 | 0.49 | 0.16 | 0.77 |
| 3000 | 0.81 | 0.17 | 0.52 | 0.2 | 0.77 |
| 3500 | 0.8 | 0.2 | 0.54 | 0.24 | 0.76 |
| 4000 | 0.79 | 0.21 | 0.56 | 0.27 | 0.75 |
| 4500 | 0.77 | 0.23 | 0.56 | 0.3 | 0.75 |
| 5000 | 0.76 | 0.23 | 0.56 | 0.3 | 0.73 |

*Figure 3: Validation: Spearman correlations for population weighted Strava heatmap outputs compared with cycle count data across different weighting buffers*

| EUROPE | Berlin | Bern | Bordeaux | Copenhagen | Glasgow | Hamburg | Helsinki | London | Lyon | Paris | Valencia | Warsaw | Zurich | average |
|---|---|---|---|---|---|---|---|---|---|---|---|---|---|---|
| raw heatmap output | 0.32 | 0.60 | 0.16 | 0.60 | 0.51 | 0.23 | 0.46 | 0.51 | 0.47 | 0.56 | 0.47 | 0.43 | 0.57 | 0.45 |
| POI buffer (m) | | | | | | | | | | | | | | |
| 100 | 0.64 | 0.48 | 0.34 | 0.63 | 0.66 | 0.28 | 0.55 | 0.67 | 0.66 | 0.49 | 0.50 | 0.57 | 0.23 | 0.51 |
| 200 | 0.70 | 0.48 | 0.64 | 0.73 | 0.75 | 0.33 | 0.63 | 0.73 | 0.75 | 0.51 | 0.61 | 0.58 | 0.38 | 0.60 |
| 500 | 0.77 | 0.62 | 0.56 | 0.76 | 0.77 | 0.46 | 0.55 | 0.80 | 0.76 | 0.59 | 0.75 | 0.59 | 0.55 | 0.66 |
| 1000 | 0.80 | 0.82 | 0.66 | 0.81 | 0.78 | 0.76 | 0.69 | 0.86 | 0.77 | 0.69 | 0.77 | 0.62 | 0.68 | 0.75 |
| 1500 | 0.80 | 0.81 | 0.70 | 0.83 | 0.79 | 0.83 | 0.78 | 0.87 | 0.79 | 0.75 | 0.75 | 0.58 | 0.74 | 0.77 |
| 2000 | 0.81 | 0.83 | 0.72 | 0.84 | 0.80 | 0.85 | 0.84 | 0.88 | 0.79 | 0.77 | 0.74 | 0.64 | 0.79 | 0.79 |
| 2500 | 0.80 | 0.81 | 0.76 | 0.85 | 0.82 | 0.86 | 0.82 | 0.89 | 0.81 | 0.77 | 0.72 | 0.69 | 0.83 | 0.80 |
| 3000 | 0.80 | 0.80 | 0.78 | 0.85 | 0.84 | 0.86 | 0.81 | 0.89 | 0.83 | 0.76 | 0.71 | 0.70 | 0.80 | 0.80 |
| 3500 | 0.80 | 0.80 | 0.78 | 0.84 | 0.85 | 0.85 | 0.82 | 0.89 | 0.84 | 0.76 | 0.68 | 0.71 | 0.77 | 0.80 |
| 4000 | 0.80 | 0.79 | 0.78 | 0.84 | 0.86 | 0.84 | 0.82 | 0.89 | 0.85 | 0.78 | 0.65 | 0.71 | 0.68 | 0.79 |
| 4500 | 0.79 | 0.76 | 0.77 | 0.83 | 0.86 | 0.83 | 0.80 | 0.89 | 0.85 | 0.79 | 0.60 | 0.74 | 0.63 | 0.78 |
| 5000 | 0.79 | 0.75 | 0.75 | 0.82 | 0.85 | 0.83 | 0.79 | 0.89 | 0.84 | 0.80 | 0.56 | 0.74 | 0.61 | 0.77 |

| NORTH AMERICA | Austin | Bakersfield | Boston | Edmonton | Los Angeles | Montreal | New York | Oakland | Philadelphia | Victoria | Washington | average |
|---|---|---|---|---|---|---|---|---|---|---|---|---|
| raw heatmap output | 0.13 | 0.31 | 0.47 | 0.46 | 0.06 | 0.27 | 0.10 | 0.25 | 0.36 | 0.27 | 0.40 | 0.28 |
| POI buffer (m) | | | | | | | | | | | | |
| 100 | 0.70 | 0.22 | 0.27 | 0.17 | 0.29 | 0.38 | 0.16 | 0.21 | 0.42 | 0.35 | 0.41 | 0.33 |
| 200 | 0.48 | 0.30 | 0.43 | 0.39 | 0.41 | 0.51 | 0.50 | 0.35 | 0.50 | 0.21 | 0.50 | 0.42 |
| 500 | 0.20 | 0.16 | 0.68 | 0.37 | 0.40 | 0.53 | 0.71 | 0.42 | 0.67 | 0.52 | 0.66 | 0.48 |
| 1000 | 0.14 | 0.31 | 0.78 | 0.36 | 0.48 | 0.64 | 0.69 | 0.60 | 0.71 | 0.55 | 0.78 | 0.55 |
| 1500 | 0.13 | 0.36 | 0.78 | 0.34 | 0.50 | 0.71 | 0.64 | 0.62 | 0.73 | 0.57 | 0.81 | 0.56 |
| 2000 | 0.20 | 0.34 | 0.78 | 0.41 | 0.48 | 0.74 | 0.67 | 0.66 | 0.75 | 0.52 | 0.80 | 0.58 |
| 2500 | 0.24 | 0.32 | 0.78 | 0.46 | 0.50 | 0.73 | 0.68 | 0.67 | 0.76 | 0.54 | 0.81 | 0.59 |
| 3000 | 0.30 | 0.31 | 0.79 | 0.55 | 0.48 | 0.73 | 0.70 | 0.71 | 0.79 | 0.53 | 0.81 | 0.61 |
| 3500 | 0.29 | 0.30 | 0.80 | 0.60 | 0.44 | 0.70 | 0.74 | 0.71 | 0.79 | 0.56 | 0.82 | 0.61 |
| 4000 | 0.30 | 0.29 | 0.81 | 0.61 | 0.43 | 0.69 | 0.74 | 0.71 | 0.80 | 0.61 | 0.81 | 0.62 |
| 4500 | 0.30 | 0.30 | 0.82 | 0.61 | 0.43 | 0.66 | 0.74 | 0.68 | 0.78 | 0.64 | 0.81 | 0.62 |
| 5000 | 0.27 | 0.30 | 0.82 | 0.62 | 0.43 | 0.67 | 0.75 | 0.67 | 0.77 | 0.63 | 0.81 | 0.61 |

| OTHER REGIONS | Buenos Aires | Christchurch | Perth | Sao Paulo | Sydney |
|---|---|---|---|---|---|
| raw heatmap output | 0.66 | 0.27 | 0.28 | 0.30 | 0.65 |
| POI buffer (m) | | | | | |
| 100 | 0.62 | 0.42 | 0.07 | 0.36 | 0.50 |
| 200 | 0.69 | 0.26 | 0.34 | 0.50 | 0.47 |
| 500 | 0.76 | 0.38 | 0.44 | 0.55 | 0.50 |
| 1000 | 0.80 | 0.29 | 0.52 | 0.59 | 0.53 |
| 1500 | 0.81 | 0.31 | 0.54 | 0.60 | 0.59 |
| 2000 | 0.82 | 0.31 | 0.56 | 0.57 | 0.63 |
| 2500 | 0.82 | 0.29 | 0.57 | 0.53 | 0.66 |
| 3000 | 0.82 | 0.28 | 0.62 | 0.50 | 0.70 |
| 3500 | 0.83 | 0.26 | 0.60 | 0.49 | 0.75 |
| 4000 | 0.83 | 0.26 | 0.60 | 0.49 | 0.78 |
| 4500 | 0.83 | 0.26 | 0.60 | 0.50 | 0.80 |
| 5000 | 0.83 | 0.25 | 0.60 | 0.50 | 0.80 |

*Figure 4: Validation: Spearman correlations for POI-weighted Strava heatmap outputs compared with cycle count data across different weighting buffers*

Regarding buffer sizes for the weighting steps, we generally observed substantially increased correlations between the weighted rasters and official counts when the weighting buffer exceeded 1000 m (Figure 5). This applied especially to cities in Europe and along the North American east coast. While the correlations were quite stable for buffers above 1000 m, the optimal buffer POI weighting differs between the cities. At a regional level, the optimal buffer size for weighting is lower in European than in North American cities. Concluding from the correlation averages, the strongest relationships were observed for 2500-3000 m weighting buffers in Europe and 3500-4000 m in North American east coast cities. Again, we found very similar patterns for population weighting and for combined pop × POI weighting (Appendix F)

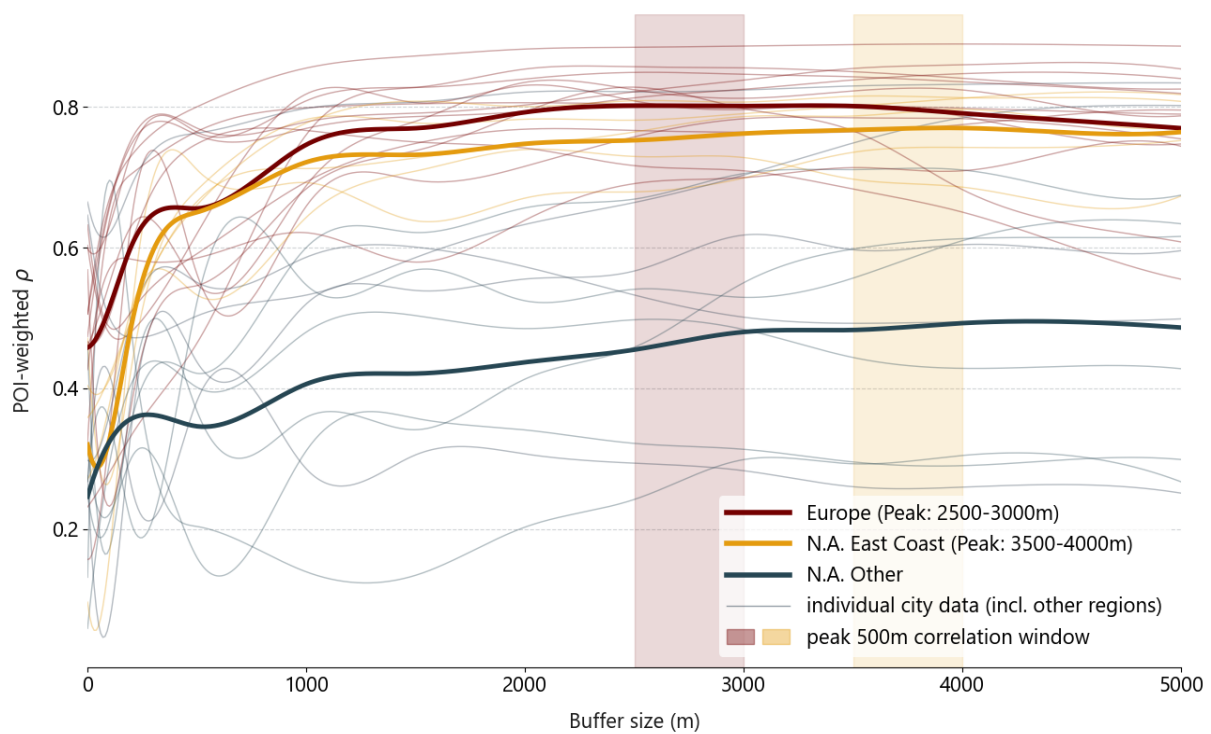

Figure 5: Observed correlations for POI-weighted Strava heatmap compared to count data plotted against the weighting buffer size. Averages are shown for Europe, North American east coast cities, and other North American cities.

Notably, the weighting did not only result in high rank correlations. Despite the Strava heatmap only indicating ordinal instead of absolute cycling counts, the weighted heatmap rasters also displayed strong alignment with the absolute counts and yielded high Pearson correlations in some cities. Again, this was especially prevalent in Europe, where POI-weighted heatmap outputs exceeded r > 0.7 in all study cities. (See Appendix D).

## 3.3 Comparison of validation outcomes with cycling mode share, population density, and count density

In addition, we tested for associations between the validation outcomes and cycling mode share, population density a count density across the study cities. Figure 6 displays the plots for POI weighting within 3000 m buffer after log-transforming the dependent variables. We found a strong correlation with r = 0.64 (p < 0.001) for cycling mode share (a), and moderate correlations with r = 0.44 (p < 0.05) for population density (b), and r = 0.48 (p < 0.01) for count density (c). Thus, our method tended to estimate cycling patterns more accurately in cities with a higher cycling modal share, and to some extent, also in cities with a higher population density. Meanwhile, a higher density of count locations yielded higher correlations in the validation outcomes. We found similar associations for population and pop × POI weighting (Appendix G).

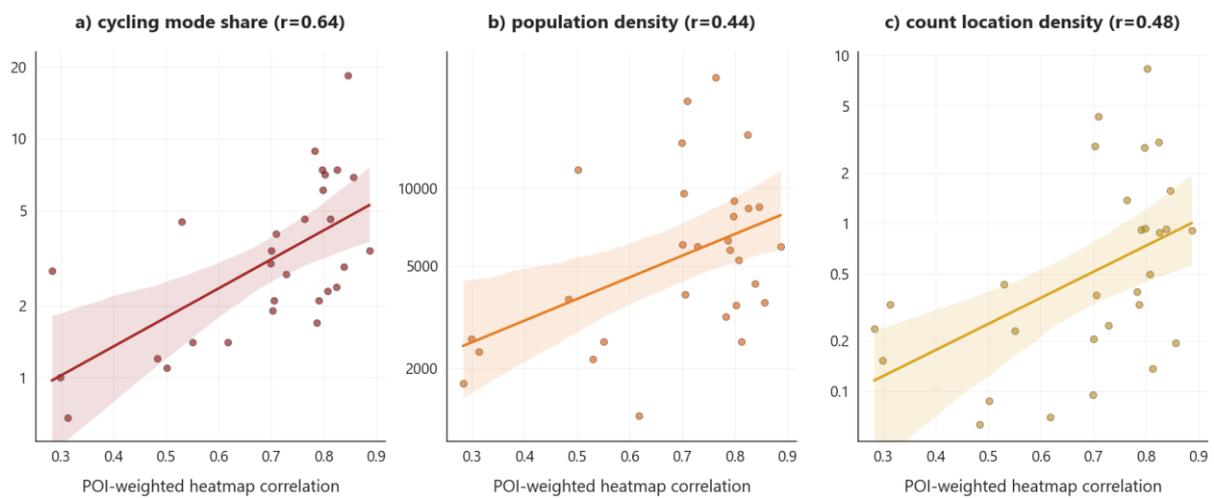

Figure 6: Validation outcomes at 3000 m POI weighting buffer plotted against (a) cycling mode share (b) population density and (c) count location density (all in log scale) in the study cities

# 4 Discussion

This study presented a global framework for estimating urban cycling patterns. We drew on the Strava Global Heatmap as an emerging data source, which we systematically assessed and evaluated. The results revealed that the heatmap, when weighted with POI counts, can reliably estimate spatial cycling intensity in many urban areas. We further showed that the performance of our proposed approach improves with the growing cycling modal share at the city level and is positively associated with the denser urban forms and higher density of cycling counters.

## 4.1 Weighting with population density and POIs

In general, our results demonstrate the importance of combining multiple data sources for cycling estimates instead of relying on either digital mobility data or land use based predictions only (Kaiser et al., 2025). By combining the Strava heatmap with a few key built environment metrics relevant to urban form, we obtained good alignment with official counts at the urban scale. A major advantage of this approach is the consistent large-scale applicability. For more fine-grained estimates of absolute cycling volumes, advanced modelling methods using various other built environment variables from city-specific datasets could likely prove even more accurate than our method (Hochmair et al., 2019; Miah et al., 2023).

To correct the internal normalization in the Strava heatmap, we found POI weighting to be more effective than population weighting, while combined weighting did not improve the accuracy compared to POI weighting alone. Primarily, these spatial weights serve to counterbalance the heatmap-specific normalization processes to make values comparable over larger geographical scales. Secondly, the underlying assumption for weighting is that these variables generally serve as a proxy for cycling intensity at a local to neighborhood scale, hence our selection of buffers. One explanation why weighting with POIs yielded particularly good alignments in many cities may be that they capture the volume and diversity of infrastructures and daily services that people most often use and cycle to. The approach presented evaluates the combination of different weighting types and buffer sizes to identify which combinations best align specifically with the Strava-derived patterns. Therefore, our results do not serve as a basis for choosing between POIs or population metrics for modeling cycling volumes in general. The previous literature has suggested that, depending on the local context and study approach, both population distribution and POI distribution can be indicative of cycling volumes (Çiriş et al., 2024; Hankey et al., 2021; Zhou et al., 2023).

Regarding appropriate buffer sizes for POI and population weighting, some variations between cities on different latitudes are expected due to the heatmap using a projected coordinate system (EPSG:3857) rather than a geographic one. Hence, depending on the latitude, the actual normalization buffer in the heatmap pre-processing can be 500-1000 m wider in North American cities compared to European ones, which could impact the results of our weighting on a similar scale. Therefore, appropriate buffer sizes for weighting can be expected to be wider in North America than in Europe. Another explanation for variations in suitable weighting buffers between European and North American cities can relate to variations in urban morphology, with generally denser cities in Europe and more dispersed ones in North America (Zhu et al., 2022). Hence, smaller weighting buffers produced higher correlations in Europe and wider buffers in North America. Therefore, any generalization of a suitable buffer size for POI weighting should be evaluated carefully. However, the results show that in cities with high correlations, the range of well-suited buffers for weighting tends to be rather broad. This suggests an appropriate buffer of 2500-3000 m in European cities, and 3500-4000 m in east coast North American cities according to our results, even if this might not be the ideal weighting buffer for single cities.

### 4.2 Potential biases of using Strava data or cycle counts

The use of Strava data for population-level cycling research has been debated because of socio-demographic biases in the usership and recorded activities (Venter et al., 2023), even if these biases may have been reduced in some cities in recent years with Strava's increasing use (Fischer et al., 2022). Additionally, as a sports tracking platform, many of the recorded cycling trips are recreational trips for exercise rather than day-to-day mobility (Lee & Sener, 2021). However, regardless of the inherent biases, our approach demonstrated that the Strava heatmap can be useful in assessing the spatial cycling patterns of urban populations at the aggregate level.

By comparing Strava heatmap outputs to cycling counters, we followed a common practice in cycling research, which is based on the assumption that counter locations can represent the ground truth of spatial cycling patterns (Boss et al., 2018; Kaiser et al., 2025). However, our validation further showed that higher correlations were found in cities with a higher density of count locations. On the other hand, cities with lower count location densities were associated with lower correlations. These results suggest that potential mismatches between Strava-derived cycling volumes and count data found in previous studies (Lee & Sener, 2021) may arise not only from spatial and other biases in the Strava data, but also from spatial and other biases in the count data. Apart from cities like London, where count locations are explicitly selected to be representative of the entire cyclable network in the city (Transport for London, 2024), cycling counts are typically collected using various methods, and vary widely from city to city in terms of density and spatiotemporal representativeness. Our observation that an increasing number and density of count locations can improve the alignment, potentially suggesting smaller amounts of spatial bias in Strava datasets than was often considered. It also provides further evidence that the presented approach can capture urban spatial cycling patterns accurately.

### 4.3 Generalizability

#### 4.3.1 Coverage

The aim of this study was to test the approach we presented for cities of varying sizes and populations with the largest possible global coverage. However, the selection of cities for the analysis was largely directed by the availability of openly accessible official cycle counts. Due to limited count availability, the study includes mainly cities from Europe and North America, with a few cities from South America and Oceania. While the study cities represent well the diversity of European and North American cities with regard to size, population, and geographic contexts, the coverage in other regions remains limited. With cycling becoming a more topical issue in cities around the world and more digital data being released from city governments, further research could attempt to verify the framework in Asian or African cities.

### 4.3.2 Applicability and data dependency

In general, the applicability of the presented framework to extract cycling patterns depends on the quality of the data sources. It relies on the POI data from OSM being mapped and tagged consistently between different cities and as comprehensively as possible to ensure generalizability. While larger scale validations and comparisons of OSM POIs are rather sparse, some existing literature indicates that OSM POIs are appropriately mapped and tagged in dense urban environments, bur declining in completeness in rural and peripheral areas (Klinkhardt et al., 2023; Mara et al., 2025; Zhang & Pfoser, 2019). The need to filter the data is also highlighted for urban planning applications (Mara et al., 2025).

Furthermore, the generalizability of the method relies on a Strava ridership that is large enough. As the number of users sharing activities increases, the aggregated spatial patterns in the heatmap align more closely with the behavior of the broader cycling population. As previous studies have demonstrated, increased ridership can directly correlate with improved representativeness; for example, the surge in Strava usage during the COVID-19 pandemic significantly enhanced the data's alignment with general cycling trends (Fischer et al., 2022). Moreover, a higher volume of cyclists in general mitigates the risk of specific subgroups (e.g., recreational mountain bikers) disproportionately skewing the observed patterns. This is reflected in our results, in which the proposed approach for deriving cycling patterns achieved higher accuracy in cities with established cycling cultures and higher modal shares.

### 4.4 Options for application in urban planning and further research

This study presents a relatively easily applicable open data approach to estimate spatial cycling patterns on an urban scale. It provides planners and policymakers with approximations on cycling path usage which can help fill data gaps when actual cycling data are sparse. However, echoing previous studies using Strava data (Jestico et al., 2016; Nelson et al., 2021), the estimated relative cycling intensities should be treated as a categorical approximation rather than directly comparable numerical values. Additionally, we emphasize that even though the validation yielded convincing results, especially in European cities, comparisons at the single segment level should be treated with care. For example, when developing urban areas, the approach presented allows planners to estimate the usage patterns of the existing cycling infrastructure at a neighborhood scale. This can help complement more traditional data sources such as counting campaigns. Additionally, the results from Strava heatmap can be combined with additional participatory datasets and local contextual understanding to assist planning for just and inclusive cycling conditions (Korpilo et al., 2025; Williams & Behrendt, 2025).

From a research perspective, the value of this study lies in the evaluation of consistent and comparable available data on cycling patterns. The presented framework can provide new options in urban cycling research, especially when data from multiple cities and countries are needed. It enables researchers to derive highly cycled and less cycled routes within cities. This could serve as a baseline for comparative analyses on cycling behavior, accessibility, or exposure, which have to date been limited due to the shortage of comparable mobility data.

## 5 Conclusions

This article presents a framework for estimating spatial urban cycling patterns and filling a crucial gap in global availability of cycling data for research. Drawing on Strava Global Heatmap and refining with population and POI data from OSM, we derived relative cycling usage intensities on an urban scale. We demonstrated that our framework achieves a strong alignment with ground truth count data, especially in Europe and along the North American east coast. Strava heatmap weighted with POI counts in 2500-3000 m buffers (Europe) or 3500-4000 m buffers (North America) correlated strongly with urban spatial cycling patterns. This adds to the evidence that crowdsourced cycling data combined with openly accessible built environment variables can be a viable source of data. It furthermore highlights the potential for self-tracked and crowdsourced data in large numbers to be applied for societal good. The presented framework can potentially be widely applied by planners and researchers alike to support decision-making for urban cycling.

# SUPPLEMENTARY MATERIAL

## A) List of excluded tags from OSM point data

'building': {'yes', 'house', 'apartments', 'residential', 'garage', 'shed'}

'natural': {'tree', 'wood', 'forest', 'water', 'coastline', 'stone', 'shrub'}

'leisure': {'park', 'garden', 'pitch', 'track'}

'place': {'suburb', 'hamlet', 'village', 'town', 'city', 'isolated_dwelling'}

'waterway': {'river', 'stream', 'canal', 'drain'}

'railway': {'switch', 'milestone', 'signal', 'crossing', 'tram_crossing', 'tram_level_crossing', 'level_crossing', 'railway_crossing', "buffer_stop"}

'amenity': {'waste_basket', 'bicycle_parking', 'grit_box', 'surveillance', 'bench'}

'man_made': {'manhole', 'surveillance'}

'highway': {'street_lamp'}

'emergency': {'fire_hydrant'}

'public_transport': {'platform'}

## B) Description and sources of cycle count data

### EUROPE

### Berlin

Verkehrsmengenkarte Rad 2023; segment-wise cycle count estimates for main street cycling network, calculated from automatic and manual strategic cycle counts

Source: Stadt Berlin, Senatsverwaltung für Mobilität, Verkehr, Klimaschutz und Umwelt ([https://www.berlin.de/sen/uvk/mobilitaet-und-verkehr/verkehrsmanagement/verkehrserhebungen/](https://www.berlin.de/sen/uvk/mobilitaet-und-verkehr/verkehrsmanagement/verkehrserhebungen/))

for validation, the counts have been attributed to the segment centroids.

### Bern

average daily counts for 2024 from automatic bike counters

Source: Verkehrsplanung Stadt Bern ([https://www.bern.ch/themen/mobilitat-und-verkehr/gesamtverkehr/basisdaten-und-erhebungen/velodaten-bern-2018-kurzbericht/bericht-velodaten-2022.pdf/view](https://www.bern.ch/themen/mobilitat-und-verkehr/gesamtverkehr/basisdaten-und-erhebungen/velodaten-bern-2018-kurzbericht/bericht-velodaten-2022.pdf/view))

### Bordeaux

total counts between 05/2024 and 04/2025 (matching heatmap) from automatic bike counters source: Bordeaux Métropole ([https://opendata.bordeaux-metropole.fr/explore/dataset/pc_captv_p_histo_heure/information/?disjunctive.gid&disjunctive.ident](https://opendata.bordeaux-metropole.fr/explore/dataset/pc_captv_p_histo_heure/information/?disjunctive.gid&disjunctive.ident))

### Copenhagen

annual average daily cycle counts 2024, based on strategic manual traffic counts

Source: Københavns Kommune ([https://www.opendata.dk/city-of-copenhagen/trafiktal](https://www.opendata.dk/city-of-copenhagen/trafiktal))

### Glasgow

total counts between 05/2024 and 04/2025 (matching heatmap) from automatic bike counters

Source: Glasgow City Council ([https://usmart.io/org/cyclingscotland/discovery/discovery-view-detail/af7fedef-7a7a-413f-a5c9-46b8eeeabeda](https://usmart.io/org/cyclingscotland/discovery/discovery-view-detail/af7fedef-7a7a-413f-a5c9-46b8eeeabeda))

**Hamburg**

total count for the week 15/05 to 21/05 2025 from automatic bike counters

Source: Stadt Hamburg, Behörde für Mobilität und Verkehrswende ([https://metaver.de/trefferanzeige?cmd=doShowDocument&docuuid=9072E37F-8505-41F0-9332-B80C02C7E802](https://metaver.de/trefferanzeige?cmd=doShowDocument&docuuid=9072E37F-8505-41F0-9332-B80C02C7E802))

**Helsinki**

total counts between 05/2024 and 04/2025 (matching heatmap) from automatic bike counters source: Helsingin seutu ([https://data.eco-counter.com/ParcPublic/?id=5589](https://data.eco-counter.com/ParcPublic/?id=5589))

**London**

manual strategic travel counts carried out in 2024. Representative for the city's cyclable network on a typical spring day in the city.

Source: Transport for London ([https://cycling.data.tfl.gov.uk/](https://cycling.data.tfl.gov.uk/))

**Lyon**

total counts between 05/2024 and 04/2025 (matching heatmap) from automatic bike counters

Source: Métropole de Lyon ([https://data.grandlyon.com/portail/fr/jeux-de-donnees/comptage-des-mobilites-de-la-metropole-de-lyon/info](https://data.grandlyon.com/portail/fr/jeux-de-donnees/comptage-des-mobilites-de-la-metropole-de-lyon/info))

**Paris**

total counts between 05/2024 and 04/2025 (matching heatmap) from automatic bike counters

Source: Ville de Paris ([https://opendata.paris.fr/explore/dataset/comptage-velo-donnees-compteurs/information/?disjunctive.id_compteur&disjunctive.nom_compteur&disjunctive.id&disjunctive.name](https://opendata.paris.fr/explore/dataset/comptage-velo-donnees-compteurs/information/?disjunctive.id_compteur&disjunctive.nom_compteur&disjunctive.id&disjunctive.name))

**Valencia**

daily average number of cyclists (available on monthly basis) from 05/2024 to 04/25 (matching heatmap) from automatic bike counters source for data and counter locations: Ajuntament de Valencia (https://www.valencia.es/cas/movilidad/otras-descargas)

**Warsaw**

total counts between 05/2024 and 04/2025 (matching heatmap) from automatic bike counters source: Zarząd Dróg Miejskich w Warszawie (https://zdm.waw.pl/dzialania/badania-i-analizy/analiza-ruchu-na-drogach/analiza-ruchu-na-drogach-online/)

**Zurich**

total counts between 05/2024 and 04/2025 (matching heatmap) from automatic bike counters

Source: Stadt Zürich (https://data.stadt-zuerich.ch/dataset/ted_taz_verkehrszaehlungen_werte_fussgaenger_velo)

**NORTH AMERICA**

**Austin**

total counts between 10/2024 and 09/2025 (matching heatmap) from automatic bike counters

Source: City of Austin (https://cityofaustin.eco-counter.com/?startDate=2024-10-01&endDate=2025-09-30&flowmode=2)

**Bakersfield**

manual one-day counts during 2024

Source: Kern Council of Governments (https://kerncog.ms2soft.com/tdms.ui/nmds/dashboard?loc=kerncog)

### Boston

one-day counts during June 2024 from automatic bike counters

Source: City of Boston (https://www.arcgis.com/apps/dashboards/3491f78e46a545fe8186ad70ebe897aa)

### Edmonton

total counts between 10/2024 and 09/2025 (matching heatmap) from automatic bike counters

Source: City of Edmonton (https://data.edmonton.ca/Monitoring-and-Data-Collection/Daily-Pedestrian-and-Bike-Counts/sw7k-ptx8/about_data)

### Los Angeles

manual one-day counts during fall 2023

Source: Los Angeles Department of Transportation (https://data.lacity.org/Transportation/2023-Walk-Bike-Count-Data/6ux4-qj74/about_data)

### Montreal

total counts between 01/2024 and 12/2024 from automatic bike counters

Source: Government and Municipalities of Québec (https://open.canada.ca/data/en/dataset/f170fecc-18db-44bc-b4fe-5b0b6d2c7297)

### New York

total counts between 10/2024 and 09/2025 (matching heatmap) from automatic bike counters

Source: New York City Department of Transportation (https://data.cityofnewyork.us/Transportation/Bicycle-Counts/uczf-rk3c/about_data)

### Oakland

one-day peak-hour counts during September 2025 from video recordings

Source: City of Oakland (https://www.oaklandca.gov/Public-Safety-Streets/Walking-and-Biking-in-Oakland/Bicycle-Planning-Evaluation)

### Philadelphia

one-day counts between 10/2024 and 09/2025 (matching heatmap) from automatic bike counters

Source: Delaware Valley Regional Planning Commission (https://catalog.dvrpc.org/dataset/bicycle-count-locations)

### Victoria

total counts between 10/2024 and 09/2025 (matching heatmap) from automatic bike counters

Source: Regional Cyclist and Pedestrian Count Program (https://data.eco-counter.com/ParcPublic/?id=4828)

### Washington

manual one-day counts during spring 2024

Source: District of Columbia, Department of Transportation (https://bikelanes.ddot.dc.gov/pages/manual-counts)

### OTHER REGIONS

### Buenos Aries

one-hour manual counts during spring and fall 2024

Source: Buenos Aires Ciudad, Secretaría de Transporte y Obras Públicas (https://data.buenosaires.gob.ar/nl/dataset/conteo-ciclistas)

### Christchurch

total count for the week 01/10 to 07/10 2025 from automatic bike counters

Source: Christchurch City Council (https://smartview.ccc.govt.nz/data/cycle-counters)

### Perth

total count of the latest year (updated June 2025) from permanent automatic counters

Source: Main Roads Western Australia (https://catalogue.data.wa.gov.au/dataset/mrwa-bike-count-sites)

**Sao Paulo**

manual one-day counts during 2023 and 2024

Source: Prefeitura de Sao Paulo ([https://www.cetsp.com.br/consultas/bicicleta/contadores-de-bicicletas.aspx](https://www.cetsp.com.br/consultas/bicicleta/contadores-de-bicicletas.aspx))

**Sydney**

manual one-day counts during October 2025

Source: City of Sydney ([https://data.cityofsydney.nsw.gov.au/datasets/5e3b0001a1a4402486d1fb76bb8b290c_0](https://data.cityofsydney.nsw.gov.au/datasets/5e3b0001a1a4402486d1fb76bb8b290c_0))

## C) Cycling mode share 2023 in the study cities

*Table 2: Cycling mode share 2023 from Google Environmental Insights Explorer*

| EUROPE | cycling modal share (%) | NORTH AMERICA | cycling modal share (%) | OTHER REGIONS | cycling modal share (%) |
|---|---|---|---|---|---|
| Bern | 7.1 | Austin | 1.0 | Buenos Aires | 2.4 |
| Berlin | 7.4 | Bakersfield | 0.68 | Christchurch | 2.8 |
| Bordeaux | 8.9 | Boston | 2.1 | Perth | 1.4 |
| Copenhagen | 18.4 | Edmonton | 1.4 | Sao Paulo | 1.1 |
| Glasgow | 2.9 | Los Angeles | 1.2 | Sydney | 1.9 |
| Hamburg | 6.9 | Montreal | 2.7 | | |
| Helsinki | 4.6 | New York | 3.0 | | |
| London | 3.4 | Oakland | 2.1 | | |
| Lyon | 7.4 | Philadelphia | 1.7 | | |
| Paris | 4.6 | Victoria | 4.5 | | |
| Valencia | 4.0 | Washington | 2.3 | | |
| Warsaw | 3.4 | | | | |
| Zurich | 6.1 | | | | |

## D) Pearson correlations for Strava heatmap outputs compared with cycle counts

| EUROPE | Berlin | Bern | Bordeaux | Copenhagen | Glasgow | Hamburg | Helsinki | London | Lyon | Paris | Valencia | Warsaw | Zurich | | average |
|---|---|---|---|---|---|---|---|---|---|---|---|---|---|---|---|
| raw heatmap output | 0.31 | 0.60 | 0.11 | 0.53 | 0.42 | 0.35 | 0.39 | 0.40 | 0.47 | 0.44 | 0.44 | 0.50 | 0.54 | | 0.42 |
| population buffer (m) | | | | | | | | | | | | | | | |
| 100 | 0.46 | 0.20 | 0.38 | 0.32 | 0.40 | 0.17 | 0.17 | 0.24 | 0.40 | 0.45 | 0.40 | 0.58 | 0.34 | | 0.35 |
| 200 | 0.54 | 0.46 | 0.45 | 0.41 | 0.47 | 0.18 | 0.13 | 0.29 | 0.54 | 0.46 | 0.54 | 0.69 | 0.62 | | 0.44 |
| 500 | 0.63 | 0.71 | 0.52 | 0.54 | 0.59 | 0.37 | 0.32 | 0.34 | 0.76 | 0.44 | 0.67 | 0.73 | 0.75 | | 0.57 |
| 1000 | 0.68 | 0.82 | 0.61 | 0.63 | 0.59 | 0.65 | 0.46 | 0.42 | 0.81 | 0.48 | 0.69 | 0.76 | 0.77 | | 0.64 |
| 1500 | 0.71 | 0.88 | 0.64 | 0.70 | 0.61 | 0.74 | 0.55 | 0.48 | 0.82 | 0.53 | 0.65 | 0.73 | 0.80 | | 0.68 |
| 2000 | 0.72 | 0.81 | 0.65 | 0.75 | 0.60 | 0.75 | 0.61 | 0.52 | 0.81 | 0.60 | 0.65 | 0.71 | 0.79 | | 0.69 |
| 2500 | 0.73 | 0.73 | 0.64 | 0.77 | 0.61 | 0.75 | 0.68 | 0.55 | 0.80 | 0.65 | 0.65 | 0.70 | 0.77 | | 0.70 |
| 3000 | 0.74 | 0.71 | 0.64 | 0.77 | 0.61 | 0.75 | 0.72 | 0.58 | 0.80 | 0.67 | 0.64 | 0.68 | 0.75 | | 0.70 |
| 3500 | 0.74 | 0.71 | 0.63 | 0.76 | 0.60 | 0.75 | 0.71 | 0.60 | 0.79 | 0.69 | 0.62 | 0.68 | 0.72 | | 0.69 |
| 4000 | 0.74 | 0.74 | 0.62 | 0.76 | 0.58 | 0.75 | 0.72 | 0.62 | 0.78 | 0.70 | 0.59 | 0.67 | 0.69 | | 0.69 |
| 4500 | 0.74 | 0.75 | 0.61 | 0.75 | 0.58 | 0.73 | 0.69 | 0.63 | 0.77 | 0.71 | 0.57 | 0.68 | 0.64 | | 0.68 |
| 5000 | 0.73 | 0.74 | 0.59 | 0.73 | 0.59 | 0.71 | 0.65 | 0.64 | 0.75 | 0.70 | 0.54 | 0.68 | 0.61 | | 0.67 |

| NORTH AMERICA | Austin | Bakersfield | Boston | Edmonton | Los Angeles | Montreal | New York | Oakland | Philadelphia | Victoria | Washington | average |
|---|---|---|---|---|---|---|---|---|---|---|---|---|
| raw heatmap output | 0.20 | 0.37 | 0.33 | 0.49 | 0.13 | 0.48 | 0.23 | 0.22 | 0.38 | 0.49 | 0.59 | 0.36 |
| population buffer (m) | | | | | | | | | | | | |
| 100 | 0.32 | 0.03 | 0.02 | 0.41 | 0.15 | 0.27 | 0.15 | -0.08 | 0.54 | 0.02 | 0.10 | 0.18 |
| 200 | 0.33 | -0.03 | 0.03 | 0.43 | 0.18 | 0.44 | 0.51 | 0.21 | 0.54 | 0.10 | 0.10 | 0.26 |
| 500 | 0.28 | -0.02 | 0.21 | 0.48 | 0.05 | 0.62 | 0.50 | 0.19 | 0.68 | 0.22 | 0.32 | 0.32 |
| 1000 | 0.25 | 0.06 | 0.37 | 0.45 | 0.00 | 0.75 | 0.23 | 0.22 | 0.58 | 0.30 | 0.43 | 0.33 |
| 1500 | 0.16 | 0.14 | 0.50 | 0.48 | -0.07 | 0.76 | 0.17 | 0.26 | 0.58 | 0.45 | 0.50 | 0.36 |
| 2000 | 0.15 | 0.20 | 0.57 | 0.55 | -0.07 | 0.76 | 0.23 | 0.32 | 0.62 | 0.57 | 0.55 | 0.40 |
| 2500 | 0.23 | 0.27 | 0.60 | 0.60 | -0.05 | 0.74 | 0.31 | 0.33 | 0.67 | 0.62 | 0.61 | 0.45 |
| 3000 | 0.30 | 0.30 | 0.58 | 0.58 | -0.03 | 0.72 | 0.31 | 0.31 | 0.70 | 0.67 | 0.66 | 0.46 |
| 3500 | 0.35 | 0.33 | 0.59 | 0.57 | -0.02 | 0.69 | 0.32 | 0.27 | 0.71 | 0.71 | 0.70 | 0.47 |
| 4000 | 0.37 | 0.37 | 0.59 | 0.56 | 0.00 | 0.68 | 0.37 | 0.25 | 0.68 | 0.72 | 0.72 | 0.48 |
| 4500 | 0.38 | 0.40 | 0.57 | 0.53 | 0.00 | 0.68 | 0.43 | 0.25 | 0.65 | 0.72 | 0.74 | 0.49 |
| 5000 | 0.38 | 0.42 | 0.57 | 0.52 | 0.02 | 0.67 | 0.45 | 0.28 | 0.62 | 0.72 | 0.74 | 0.49 |

| OTHER REGIONS | Buenos Aires | Christchurch | Perth | Sao Paulo | Sydney |
|---|---|---|---|---|---|
| raw heatmap output | 0.61 | 0.26 | 0.29 | 0.28 | 0.55 |
| population buffer (m) | | | | | |
| 100 | 0.66 | -0.20 | 0.15 | 0.07 | 0.29 |
| 200 | 0.68 | -0.17 | 0.15 | 0.07 | 0.30 |
| 500 | 0.72 | -0.30 | 0.17 | 0.15 | 0.37 |
| 1000 | 0.74 | -0.28 | 0.24 | 0.23 | 0.50 |
| 1500 | 0.75 | -0.14 | 0.32 | 0.24 | 0.55 |
| 2000 | 0.75 | 0.03 | 0.40 | 0.24 | 0.58 |
| 2500 | 0.75 | 0.18 | 0.47 | 0.24 | 0.61 |
| 3000 | 0.74 | 0.28 | 0.51 | 0.26 | 0.61 |
| 3500 | 0.72 | 0.34 | 0.54 | 0.28 | 0.59 |
| 4000 | 0.71 | 0.36 | 0.55 | 0.27 | 0.58 |
| 4500 | 0.69 | 0.38 | 0.54 | 0.28 | 0.58 |
| 5000 | 0.67 | 0.38 | 0.55 | 0.27 | 0.57 |

*Figure 1: Validation: Spearman correlations for population weighted Strava heatmap outputs compared with cycle count data across different weighting buffers*

| EUROPE | Berlin | Bern | Bordeaux | Copenhagen | Glasgow | Hamburg | Helsinki | London | Lyon | Paris | Valencia | Warsaw | Zurich | | average |
|---|---|---|---|---|---|---|---|---|---|---|---|---|---|---|---|
| raw heatmap output | 0.31 | 0.60 | 0.11 | 0.53 | 0.42 | 0.35 | 0.39 | 0.40 | 0.47 | 0.44 | 0.44 | 0.50 | 0.54 | | 0.42 |
| POI buffer (m) | | | | | | | | | | | | | | | |
| 100 | 0.58 | 0.54 | 0.34 | 0.45 | 0.64 | -0.06 | 0.27 | 0.49 | 0.64 | 0.76 | 0.68 | 0.43 | 0.22 | | 0.46 |
| 200 | 0.64 | 0.52 | 0.42 | 0.54 | 0.59 | 0.02 | 0.37 | 0.56 | 0.60 | 0.74 | 0.76 | 0.51 | 0.44 | | 0.52 |
| 500 | 0.72 | 0.74 | 0.55 | 0.65 | 0.69 | 0.25 | 0.35 | 0.65 | 0.71 | 0.82 | 0.83 | 0.54 | 0.53 | | 0.62 |
| 1000 | 0.77 | 0.86 | 0.65 | 0.77 | 0.73 | 0.65 | 0.45 | 0.71 | 0.75 | 0.84 | 0.83 | 0.53 | 0.67 | | 0.71 |
| 1500 | 0.79 | 0.83 | 0.70 | 0.80 | 0.77 | 0.75 | 0.55 | 0.75 | 0.79 | 0.85 | 0.76 | 0.53 | 0.76 | | 0.74 |
| 2000 | 0.80 | 0.80 | 0.73 | 0.79 | 0.78 | 0.78 | 0.62 | 0.77 | 0.80 | 0.84 | 0.72 | 0.54 | 0.80 | | 0.75 |
| 2500 | 0.80 | 0.75 | 0.72 | 0.80 | 0.79 | 0.78 | 0.68 | 0.78 | 0.80 | 0.83 | 0.70 | 0.59 | 0.79 | | 0.75 |
| 3000 | 0.80 | 0.75 | 0.70 | 0.79 | 0.79 | 0.77 | 0.74 | 0.78 | 0.80 | 0.79 | 0.68 | 0.60 | 0.76 | | 0.75 |
| 3500 | 0.80 | 0.71 | 0.67 | 0.77 | 0.79 | 0.78 | 0.76 | 0.78 | 0.80 | 0.77 | 0.64 | 0.62 | 0.71 | | 0.74 |
| 4000 | 0.80 | 0.71 | 0.65 | 0.76 | 0.78 | 0.78 | 0.78 | 0.77 | 0.79 | 0.76 | 0.60 | 0.64 | 0.66 | | 0.73 |
| 4500 | 0.80 | 0.68 | 0.63 | 0.75 | 0.76 | 0.78 | 0.76 | 0.77 | 0.77 | 0.76 | 0.56 | 0.69 | 0.62 | | 0.72 |
| 5000 | 0.80 | 0.71 | 0.60 | 0.74 | 0.74 | 0.77 | 0.73 | 0.77 | 0.76 | 0.74 | 0.52 | 0.70 | 0.58 | | 0.70 |

| NORTH AMERICA | Austin | Bakersfield | Boston | Edmonton | Los Angeles | Montreal | New York | Oakland | Philadelphia | Victoria | Washington | average |
|---|---|---|---|---|---|---|---|---|---|---|---|---|
| raw heatmap output | 0.20 | 0.37 | 0.33 | 0.49 | 0.13 | 0.48 | 0.23 | 0.22 | 0.38 | 0.49 | 0.59 | 0.36 |
| POI buffer (m) | | | | | | | | | | | | |
| 100 | 0.67 | 0.05 | 0.09 | 0.24 | 0.10 | 0.33 | 0.30 | 0.01 | 0.07 | 0.36 | 0.27 | 0.23 |
| 200 | 0.52 | 0.03 | 0.19 | 0.41 | 0.18 | 0.41 | 0.51 | 0.06 | 0.03 | 0.51 | 0.33 | 0.29 |
| 500 | 0.48 | -0.05 | 0.34 | 0.33 | 0.23 | 0.41 | 0.66 | 0.22 | 0.53 | 0.55 | 0.47 | 0.38 |
| 1000 | 0.49 | 0.15 | 0.49 | 0.36 | 0.33 | 0.53 | 0.64 | 0.43 | 0.71 | 0.58 | 0.54 | 0.48 |
| 1500 | 0.49 | 0.20 | 0.58 | 0.43 | 0.35 | 0.63 | 0.61 | 0.51 | 0.74 | 0.62 | 0.61 | 0.52 |
| 2000 | 0.51 | 0.58 | 0.63 | 0.51 | 0.35 | 0.66 | 0.61 | 0.63 | 0.71 | 0.63 | 0.65 | 0.59 |
| 2500 | 0.55 | 0.51 | 0.65 | 0.54 | 0.34 | 0.67 | 0.62 | 0.68 | 0.71 | 0.59 | 0.66 | 0.59 |
| 3000 | 0.58 | 0.49 | 0.65 | 0.61 | 0.32 | 0.66 | 0.65 | 0.69 | 0.71 | 0.56 | 0.67 | 0.60 |
| 3500 | 0.58 | 0.36 | 0.66 | 0.64 | 0.29 | 0.65 | 0.70 | 0.69 | 0.72 | 0.57 | 0.68 | 0.60 |
| 4000 | 0.58 | 0.17 | 0.67 | 0.59 | 0.28 | 0.63 | 0.74 | 0.67 | 0.73 | 0.62 | 0.71 | 0.58 |
| 4500 | 0.56 | 0.05 | 0.69 | 0.57 | 0.27 | 0.61 | 0.76 | 0.64 | 0.74 | 0.65 | 0.72 | 0.57 |
| 5000 | 0.54 | -0.01 | 0.69 | 0.57 | 0.28 | 0.61 | 0.79 | 0.61 | 0.74 | 0.65 | 0.73 | 0.56 |

| OTHER REGIONS | Buenos Aires | Christchurch | Perth | Sao Paulo | Sydney |
|---|---|---|---|---|---|
| raw heatmap output | 0.61 | 0.26 | 0.29 | 0.28 | 0.55 |
| POI buffer (m) | | | | | |
| 100 | 0.46 | 0.22 | 0.04 | 0.65 | 0.43 |
| 200 | 0.50 | 0.16 | 0.21 | 0.70 | 0.38 |
| 500 | 0.55 | 0.58 | 0.35 | 0.70 | 0.39 |
| 1000 | 0.61 | 0.63 | 0.35 | 0.68 | 0.46 |
| 1500 | 0.65 | 0.62 | 0.27 | 0.65 | 0.55 |
| 2000 | 0.69 | 0.55 | 0.27 | 0.63 | 0.58 |
| 2500 | 0.72 | 0.52 | 0.32 | 0.59 | 0.62 |
| 3000 | 0.74 | 0.48 | 0.39 | 0.56 | 0.65 |
| 3500 | 0.75 | 0.43 | 0.48 | 0.53 | 0.68 |
| 4000 | 0.76 | 0.42 | 0.53 | 0.52 | 0.69 |
| 4500 | 0.77 | 0.41 | 0.55 | 0.52 | 0.69 |
| 5000 | 0.78 | 0.38 | 0.56 | 0.53 | 0.68 |

*Figure 2: Validation: Pearson correlations for POI-weighted Strava heatmap outputs compared with cycle count data across different weighting buffers*

| EUROPE | Berlin | Bern | Bordeaux | Copenhagen | Glasgow | Hamburg | Helsinki | London | Lyon | Paris | Valencia | Warsaw | Zurich | | average |
|---|---|---|---|---|---|---|---|---|---|---|---|---|---|---|---|
| raw heatmap output | 0.30 | 0.60 | 0.17 | 0.53 | 0.43 | 0.35 | 0.40 | 0.40 | 0.42 | 0.44 | 0.44 | 0.50 | 0.54 | | 0.43 |
| pop x POI buffer (m) | | | | | | | | | | | | | | | |
| 100 | 0.60 | 0.45 | 0.38 | 0.43 | 0.42 | 0.12 | 0.33 | 0.42 | 0.51 | 0.69 | 0.68 | 0.57 | 0.38 | | 0.46 |
| 200 | 0.66 | 0.58 | 0.46 | 0.56 | 0.47 | 0.16 | 0.31 | 0.48 | 0.61 | 0.71 | 0.78 | 0.64 | 0.55 | | 0.54 |
| 500 | 0.72 | 0.76 | 0.57 | 0.67 | 0.51 | 0.37 | 0.38 | 0.57 | 0.74 | 0.77 | 0.84 | 0.65 | 0.65 | | 0.63 |
| 1000 | 0.76 | 0.85 | 0.66 | 0.77 | 0.48 | 0.69 | 0.49 | 0.66 | 0.77 | 0.78 | 0.84 | 0.65 | 0.73 | | 0.70 |
| 1500 | 0.78 | 0.86 | 0.70 | 0.82 | 0.50 | 0.76 | 0.57 | 0.71 | 0.79 | 0.79 | 0.77 | 0.63 | 0.79 | | 0.73 |
| 2000 | 0.79 | 0.81 | 0.71 | 0.81 | 0.53 | 0.78 | 0.63 | 0.74 | 0.78 | 0.81 | 0.73 | 0.62 | 0.80 | | 0.73 |
| 2500 | 0.79 | 0.75 | 0.69 | 0.81 | 0.60 | 0.77 | 0.70 | 0.75 | 0.78 | 0.80 | 0.72 | 0.64 | 0.79 | | 0.74 |
| 3000 | 0.79 | 0.74 | 0.68 | 0.79 | 0.63 | 0.77 | 0.74 | 0.75 | 0.77 | 0.77 | 0.70 | 0.63 | 0.76 | | 0.73 |
| 3500 | 0.79 | 0.73 | 0.66 | 0.78 | 0.64 | 0.77 | 0.76 | 0.75 | 0.76 | 0.74 | 0.67 | 0.64 | 0.73 | | 0.72 |
| 4000 | 0.79 | 0.74 | 0.65 | 0.77 | 0.62 | 0.77 | 0.77 | 0.74 | 0.76 | 0.74 | 0.63 | 0.65 | 0.68 | | 0.71 |
| 4500 | 0.79 | 0.73 | 0.63 | 0.76 | 0.63 | 0.76 | 0.74 | 0.74 | 0.74 | 0.72 | 0.58 | 0.66 | 0.64 | | 0.70 |
| 5000 | 0.78 | 0.73 | 0.61 | 0.74 | 0.62 | 0.75 | 0.71 | 0.74 | 0.72 | 0.70 | 0.54 | 0.67 | 0.60 | | 0.69 |

| NORTH AMERICA | Austin | Bakersfield | Boston | Edmonton | Los Angeles | Montreal | New York | Oakland | Philadelphia | Victoria | Washington | | average |
|---|---|---|---|---|---|---|---|---|---|---|---|---|---|
| raw heatmap output | 0.202 | 0.368 | 0.333 | 0.487 | 0.13 | 0.476 | 0.233 | 0.016 | 0.382 | 0.488 | 0.589 | | 0.34 |
| pop x POI buffer (m) | | | | | | | | | | | | | |
| 100 | 0.48 | 0.04 | 0.08 | 0.40 | 0.23 | 0.33 | 0.32 | 0.03 | 0.46 | 0.22 | 0.23 | | 0.26 |
| 200 | 0.54 | 0.05 | 0.14 | 0.51 | 0.25 | 0.51 | 0.62 | 0.25 | 0.53 | 0.54 | 0.23 | | 0.38 |
| 500 | 0.58 | 0.08 | 0.33 | 0.51 | 0.29 | 0.61 | 0.66 | 0.28 | 0.76 | 0.59 | 0.47 | | 0.47 |
| 1000 | 0.50 | 0.15 | 0.53 | 0.46 | 0.32 | 0.77 | 0.60 | 0.34 | 0.75 | 0.57 | 0.55 | | 0.50 |
| 1500 | 0.49 | 0.33 | 0.63 | 0.51 | 0.29 | 0.80 | 0.61 | 0.39 | 0.75 | 0.59 | 0.62 | | 0.55 |
| 2000 | 0.49 | 0.35 | 0.65 | 0.58 | 0.26 | 0.80 | 0.67 | 0.46 | 0.76 | 0.58 | 0.67 | | 0.57 |
| 2500 | 0.51 | 0.36 | 0.67 | 0.63 | 0.24 | 0.79 | 0.75 | 0.49 | 0.77 | 0.60 | 0.71 | | 0.59 |
| 3000 | 0.50 | 0.30 | 0.66 | 0.64 | 0.21 | 0.76 | 0.79 | 0.49 | 0.78 | 0.63 | 0.75 | | 0.59 |
| 3500 | 0.50 | 0.25 | 0.66 | 0.61 | 0.21 | 0.74 | 0.81 | 0.47 | 0.78 | 0.67 | 0.79 | | 0.59 |
| 4000 | 0.47 | 0.21 | 0.64 | 0.59 | 0.21 | 0.73 | 0.82 | 0.43 | 0.76 | 0.69 | 0.82 | | 0.58 |
| 4500 | 0.46 | 0.19 | 0.63 | 0.59 | 0.20 | 0.73 | 0.83 | 0.39 | 0.75 | 0.70 | 0.82 | | 0.57 |
| 5000 | 0.45 | 0.18 | 0.62 | 0.59 | 0.20 | 0.73 | 0.83 | 0.37 | 0.72 | 0.69 | 0.81 | | 0.56 |

| OTHER REGIONS | Buenos Aires | Christchurch | Perth | Sao Paulo | Sydney |
|---|---|---|---|---|---|
| raw heatmap output | 0.61 | 0.26 | 0.29 | 0.28 | 0.55 |
| pop x POI buffer (m) | | | | | |
| 100 | 0.61 | 0.08 | 0.21 | 0.42 | 0.40 |
| 200 | 0.66 | 0.02 | 0.41 | 0.58 | 0.36 |
| 500 | 0.69 | 0.09 | 0.39 | 0.61 | 0.43 |
| 1000 | 0.73 | 0.19 | 0.37 | 0.60 | 0.55 |
| 1500 | 0.76 | 0.35 | 0.41 | 0.56 | 0.59 |
| 2000 | 0.78 | 0.39 | 0.51 | 0.53 | 0.63 |
| 2500 | 0.78 | 0.42 | 0.58 | 0.50 | 0.66 |
| 3000 | 0.78 | 0.42 | 0.61 | 0.51 | 0.68 |
| 3500 | 0.77 | 0.42 | 0.65 | 0.52 | 0.67 |
| 4000 | 0.76 | 0.42 | 0.67 | 0.51 | 0.65 |
| 4500 | 0.75 | 0.41 | 0.69 | 0.51 | 0.64 |
| 5000 | 0.73 | 0.40 | 0.69 | 0.50 | 0.61 |

Figure 3: Validation: Pearson correlations for pop x POI weighted Strava heatmap outputs compared with cycle count data across different weighting buffers

## E) Spearman rank correlations for pop x POI weighted Strava heatmap outputs compared with cycle counts

**EUROPE**

| | Berlin | Bern | Bordeaux | Copenhagen | Glasgow | Hamburg | Helsinki | London | Lyon | Paris | Valencia | Warsaw | Zurich | average |
|---|---|---|---|---|---|---|---|---|---|---|---|---|---|---|
| raw heatmap output | 0.32 | 0.60 | 0.16 | 0.60 | 0.51 | 0.23 | 0.46 | 0.51 | 0.47 | 0.56 | 0.47 | 0.43 | 0.57 | 0.45 |
| pop x POI buffer (m) | | | | | | | | | | | | | | |
| 100 | 0.64 | 0.43 | 0.37 | 0.57 | 0.38 | 0.26 | 0.52 | 0.64 | 0.67 | 0.50 | 0.55 | 0.60 | 0.30 | 0.49 |
| 200 | 0.70 | 0.53 | 0.58 | 0.69 | 0.46 | 0.31 | 0.47 | 0.71 | 0.77 | 0.49 | 0.64 | 0.66 | 0.42 | 0.57 |
| 500 | 0.76 | 0.66 | 0.55 | 0.76 | 0.58 | 0.52 | 0.47 | 0.78 | 0.80 | 0.53 | 0.77 | 0.70 | 0.62 | 0.65 |
| 1000 | 0.79 | 0.81 | 0.68 | 0.81 | 0.68 | 0.80 | 0.61 | 0.84 | 0.83 | 0.61 | 0.79 | 0.75 | 0.70 | 0.75 |
| 1500 | 0.80 | 0.84 | 0.72 | 0.84 | 0.72 | 0.84 | 0.71 | 0.86 | 0.83 | 0.69 | 0.77 | 0.69 | 0.77 | 0.78 |
| 2000 | 0.80 | 0.84 | 0.73 | 0.84 | 0.77 | 0.86 | 0.80 | 0.86 | 0.84 | 0.74 | 0.75 | 0.69 | 0.83 | 0.80 |
| 2500 | 0.80 | 0.81 | 0.76 | 0.84 | 0.82 | 0.86 | 0.82 | 0.87 | 0.85 | 0.77 | 0.72 | 0.72 | 0.83 | 0.80 |
| 3000 | 0.80 | 0.81 | 0.77 | 0.84 | 0.84 | 0.86 | 0.82 | 0.87 | 0.86 | 0.77 | 0.70 | 0.71 | 0.81 | 0.80 |
| 3500 | 0.80 | 0.78 | 0.77 | 0.83 | 0.84 | 0.84 | 0.81 | 0.87 | 0.87 | 0.76 | 0.68 | 0.70 | 0.79 | 0.79 |
| 4000 | 0.80 | 0.77 | 0.76 | 0.83 | 0.83 | 0.83 | 0.79 | 0.87 | 0.87 | 0.76 | 0.66 | 0.72 | 0.72 | 0.79 |
| 4500 | 0.79 | 0.76 | 0.74 | 0.81 | 0.83 | 0.82 | 0.78 | 0.87 | 0.87 | 0.75 | 0.62 | 0.72 | 0.69 | 0.77 |
| 5000 | 0.79 | 0.75 | 0.73 | 0.80 | 0.81 | 0.81 | 0.76 | 0.87 | 0.86 | 0.75 | 0.57 | 0.72 | 0.67 | 0.76 |

**NORTH AMERICA**

| | Austin | Bakersfield | Boston | Edmonton | Los Angeles | Montreal | New York | Oakland | Philadelphia | Victoria | Washington | average |
|---|---|---|---|---|---|---|---|---|---|---|---|---|
| raw heatmap output | 0.13 | 0.31 | 0.47 | 0.46 | 0.06 | 0.27 | 0.10 | 0.25 | 0.36 | 0.27 | 0.40 | 0.28 |
| pop x POI buffer (m) | | | | | | | | | | | | |
| 100 | 0.13 | 0.26 | 0.02 | 0.24 | 0.29 | 0.38 | 0.46 | 0.15 | 0.64 | 0.18 | 0.36 | 0.28 |
| 200 | 0.21 | 0.37 | 0.30 | 0.40 | 0.29 | 0.60 | 0.72 | 0.31 | 0.63 | 0.33 | 0.33 | 0.41 |
| 500 | 0.10 | 0.34 | 0.65 | 0.41 | 0.44 | 0.67 | 0.74 | 0.38 | 0.71 | 0.60 | 0.61 | 0.51 |
| 1000 | 0.15 | 0.37 | 0.75 | 0.34 | 0.47 | 0.80 | 0.65 | 0.50 | 0.71 | 0.60 | 0.71 | 0.55 |
| 1500 | 0.29 | 0.37 | 0.77 | 0.41 | 0.43 | 0.84 | 0.66 | 0.51 | 0.73 | 0.59 | 0.78 | 0.58 |
| 2000 | 0.37 | 0.35 | 0.78 | 0.48 | 0.42 | 0.83 | 0.76 | 0.50 | 0.77 | 0.57 | 0.80 | 0.60 |
| 2500 | 0.40 | 0.36 | 0.79 | 0.56 | 0.42 | 0.83 | 0.81 | 0.53 | 0.78 | 0.59 | 0.81 | 0.63 |
| 3000 | 0.38 | 0.34 | 0.80 | 0.61 | 0.39 | 0.80 | 0.82 | 0.54 | 0.77 | 0.61 | 0.83 | 0.63 |
| 3500 | 0.40 | 0.33 | 0.80 | 0.61 | 0.40 | 0.77 | 0.82 | 0.57 | 0.75 | 0.67 | 0.84 | 0.63 |
| 4000 | 0.40 | 0.33 | 0.79 | 0.61 | 0.39 | 0.77 | 0.81 | 0.53 | 0.75 | 0.72 | 0.85 | 0.63 |
| 4500 | 0.43 | 0.33 | 0.78 | 0.61 | 0.39 | 0.78 | 0.83 | 0.49 | 0.72 | 0.73 | 0.85 | 0.63 |
| 5000 | 0.44 | 0.33 | 0.78 | 0.61 | 0.38 | 0.78 | 0.82 | 0.48 | 0.70 | 0.75 | 0.82 | 0.63 |

**OTHER REGIONS**

| | Buenos Aires | Christchurch | Perth | Sao Paulo | Sydney |
|---|---|---|---|---|---|
| raw heatmap output | 0.66 | 0.27 | 0.28 | 0.30 | 0.65 |
| pop x POI buffer (m) | | | | | |
| 100 | 0.71 | 0.23 | 0.17 | 0.37 | 0.40 |
| 200 | 0.81 | 0.07 | 0.41 | 0.40 | 0.48 |
| 500 | 0.84 | 0.09 | 0.39 | 0.49 | 0.57 |
| 1000 | 0.85 | 0.11 | 0.57 | 0.57 | 0.61 |
| 1500 | 0.85 | 0.20 | 0.60 | 0.56 | 0.67 |
| 2000 | 0.85 | 0.24 | 0.61 | 0.52 | 0.73 |
| 2500 | 0.85 | 0.24 | 0.63 | 0.52 | 0.78 |
| 3000 | 0.85 | 0.24 | 0.64 | 0.51 | 0.81 |
| 3500 | 0.84 | 0.24 | 0.65 | 0.53 | 0.81 |
| 4000 | 0.84 | 0.25 | 0.66 | 0.54 | 0.81 |
| 4500 | 0.83 | 0.26 | 0.68 | 0.57 | 0.80 |
| 5000 | 0.81 | 0.25 | 0.69 | 0.58 | 0.78 |

*Figure 7: Validation: Spearman correlations for pop x POI weighted Strava heatmap outputs compared with cycle count data across different weighting buffers*

## F) Distribution of Spearman correlations for varying buffer sizes

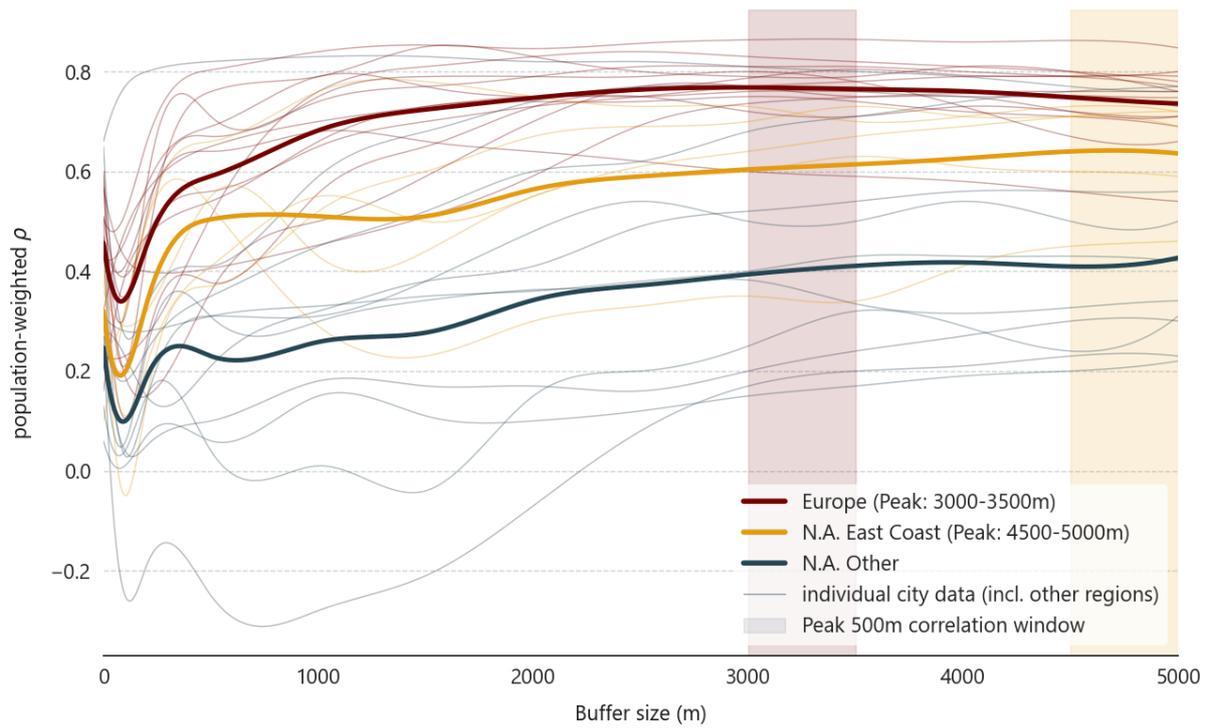

Figure 8: Observed correlations for population weighted Strava heatmap compared to count data plotted against the weighting buffer size. Averages are shown for Europe, North American east coast cities, and other North American cities.

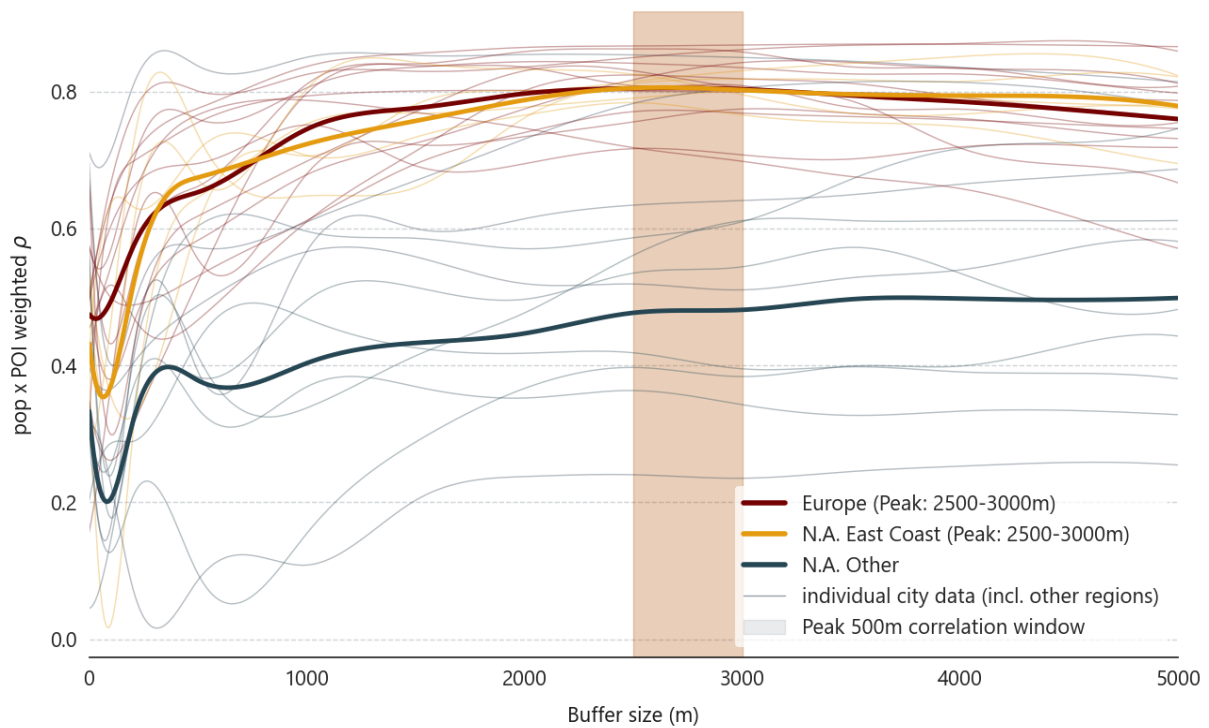

Figure 6: Observed correlations for pop x POI weighted Strava heatmap compared to count data plotted against the weighting buffer size. Averages are shown for Europe, North American east coast cities, and other North American cities.

## G) Comparison of the heatmap outputs with cycling mode share, population density, and count location density

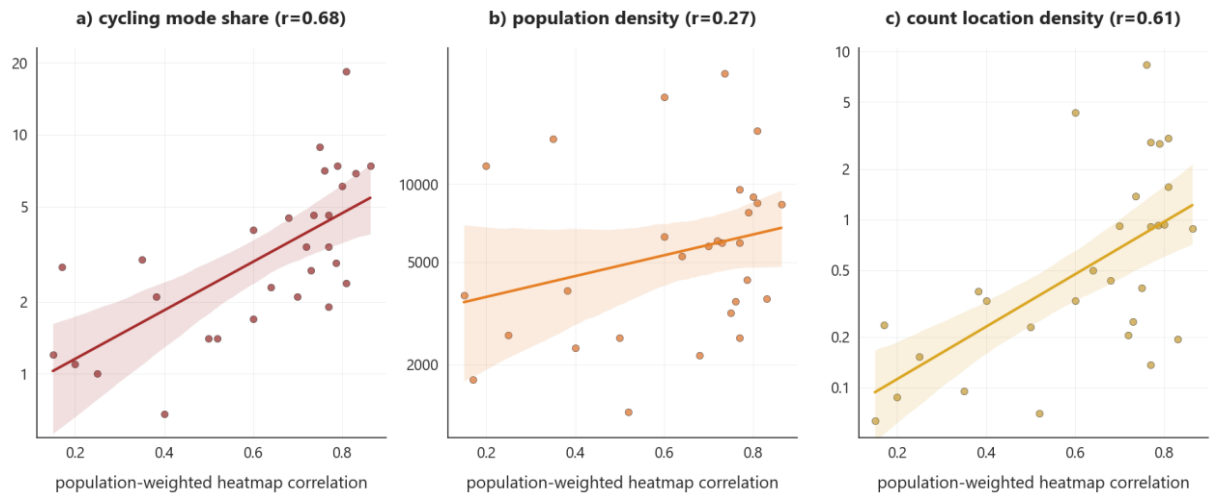

Figure 7: Validation outcomes at 3000 m population weighting buffer plotted against (a) cycling mode share (b) population density and (c) count location density (all in log scale) in the study cities

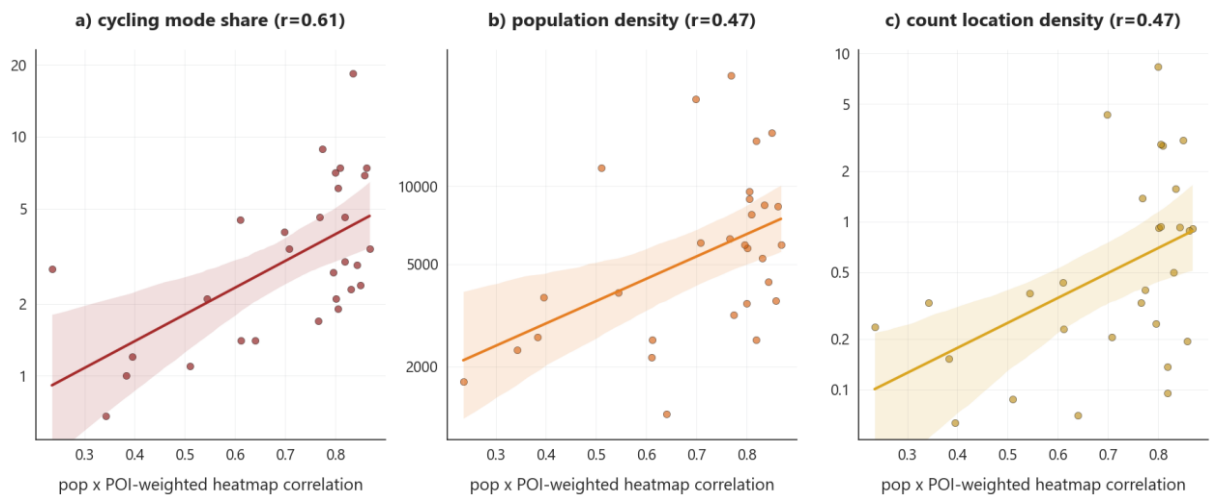

Figure 8: Validation outcomes at 3000 m pop x POI weighting buffer plotted against (a) cycling mode share (b) population density and (c) count location density (all in log scale) in the study cities